\documentclass[11pt]{article}

\usepackage{aas_macros,amsmath,amssymb,cite,esint,graphicx,mathtools,transparent}
\usepackage[margin=1in,letterpaper]{geometry}
\usepackage[colorlinks=true]{hyperref}
\usepackage[affil-it]{authblk}
\usepackage[normalem]{ulem}

\numberwithin{equation}{section}
\let\originalleft\left
\let\originalright\right
\renewcommand{\left}{\mathopen{}\mathclose\bgroup\originalleft}
\renewcommand{\right}{\aftergroup\egroup\originalright}

\newcommand{\ab}[1]{\left|#1\right|}
\newcommand{\av}[1]{\left\langle#1\right\rangle}
\newcommand{\br}[1]{\left[#1\right]}
\newcommand{\cu}[1]{\left\{#1\right\}}
\newcommand{\pa}[1]{\left(#1\right)}
\renewcommand{\O}[1]{\mathcal{O}\pa{#1}}

\newcommand{\dt}{\mathop{}\!\delta}
\newcommand{\ed}{\mathop{}\!\mathrm{d}}
\newcommand{\pd}{\mathop{}\!\partial}
\newcommand{\tr}{\tilde{r}}

\DeclareMathOperator\arctanh{arctanh}
\DeclareMathOperator\erf{erf}
\DeclareMathOperator\sign{sign}
\DeclareMathOperator\sn{sn}

\begin{document}

\title{\Huge Photon Ring Autocorrelations\vspace{-0.25cm}\\
{\transparent{0.75}\Large Photon Ring Autocorrelations}\vspace{-0.5cm}\\
{\transparent{0.5}\small Photon Ring Autocorrelations}\vspace{-0.75cm}\\
{\transparent{0.25}\tiny Photon Ring Autocorrelations}}

\date{}
\author[1]{Shahar~Hadar\thanks{shaharhadar@g.harvard.edu}}
\author[2]{Michael~D.~Johnson\thanks{mjohnson@cfa.harvard.edu}}
\author[3]{Alexandru~Lupsasca\thanks{lupsasca@princeton.edu}}
\author[4]{George~N.~Wong\thanks{gnwong2@illinois.edu}}
\affil[1]{\small
Center for the Fundamental Laws of Nature, Harvard University, Cambridge, MA 02138, USA}
\affil[2]{\small
Center for Astrophysics $|$ Harvard \& Smithsonian, Cambridge, MA 02138, USA}
\affil[3]{\small
Princeton Gravity Initiative, Princeton University, Princeton, NJ 08544, USA}
\affil[4]{\small
Department of Physics, University of Illinois, Urbana, IL 61801, USA}

\maketitle

\begin{abstract}
In the presence of a black hole, light sources connect to observers along multiple paths.  As a result, observed brightness fluctuations must be correlated across different times and positions in black hole images.  Photons that execute multiple orbits around the black hole appear near a critical curve in the observer sky, giving rise to the photon ring.  In this paper, a novel observable is proposed: the two-point correlation function of intensity fluctuations on the photon ring.  This correlation function is analytically computed for a Kerr black hole surrounded by stochastic equatorial emission, with source statistics motivated by simulations of a turbulent accretion flow.  It is shown that this two-point function exhibits a universal, self-similar structure consisting of multiple peaks of identical shape: while the profile of each peak encodes statistical properties of fluctuations in the source, the locations and heights of the peaks are determined purely by the black hole parameters.  Measuring these peaks would demonstrate the existence of the photon ring without resolving its thickness, and would provide estimates of black hole mass and spin.  With regular monitoring over sufficiently long timescales, this measurement could be possible via interferometric imaging with modest improvements to the Event Horizon Telescope.
\end{abstract}

\clearpage

\section{Introduction}
\label{sec:Introduction}

The remarkable first image of a black hole (BH), obtained last year by the Event Horizon Telescope (EHT) Collaboration using Very-Long-Baseline Interferometry (VLBI) \cite{EHT2019a,EHT2019b,EHT2019c,EHT2019d,EHT2019e,EHT2019f}, was the culmination of a decades-long effort to peer deeper into the neighboring galaxy M87 and image the supermassive black hole M87* at its center.  This breakthrough marks the start of a new era in which we expect to obtain progressively better images of increasingly many BHs.  It is therefore of great interest to theorists to provide detailed predictions:  What do we expect to see in BH images?  How can we distinguish features of these images that depend on the complex astrophysical environment of the BH from universal properties that depend only on the BH itself?  And how can we use these universal features to test General Relativity and measure the mass and spin of the BH?

The propagation of light around a BH was first studied a century ago \cite{Hilbert1917}, immediately following the discovery of the Schwarzschild metric.  Basic features of BH images and their dependence on BH spin have been investigated over the ensuing decades \cite{Darwin1959,Bardeen1973,Luminet1979,Falcke2000,Beckwith2005}.  In recent years, increasingly complex simulations have produced sophisticated models of the astrophysical environment around a BH, enabling detailed numerical studies of BH images \cite{Gammie2003,EHT2019e,Porth2019}.  The past year has witnessed a flurry of activity in the analytic study of the time-averaged image $\av{I(\rho,\varphi)}$ of a BH, where $I$ is the specific intensity at polar coordinates $(\rho,\varphi)$ centered about the BH's position in the observer sky \cite{Gralla2019,Johnson2020,GrallaLupsasca2020a,Narayan2019,Gralla2020,Farah2020,GrallaLupsasca2020c,GrallaLupsasca2020d}.  When a BH is surrounded by optically thin emitting material, its image displays a narrow ring of enhanced brightness: the \textit{photon ring}.  This ring is a (formally infinite) sum of increasingly demagnified subrings, each a strongly lensed image of the direct emission.  These subrings asymptote to a critical curve in the observer sky, first derived by Bardeen \cite{Bardeen1973}.  The demagnification factor at every angle in the sky is related to properties of the \textit{photon shell}, a region of spacetime in the vicinity of the BH that admits (unstable) bound photon orbits; more specifically, to the Lyapunov exponent $\gamma$ characterizing the orbital instability of these geodesics \cite{Johnson2020}.  Two additional critical exponents of photon shell orbits, $\tau$ and $\delta$, characterize their temporal and azimuthal periods, respectively \cite{GrallaLupsasca2020a}.

\begin{figure}
	\centering
	\includegraphics[width=\textwidth]{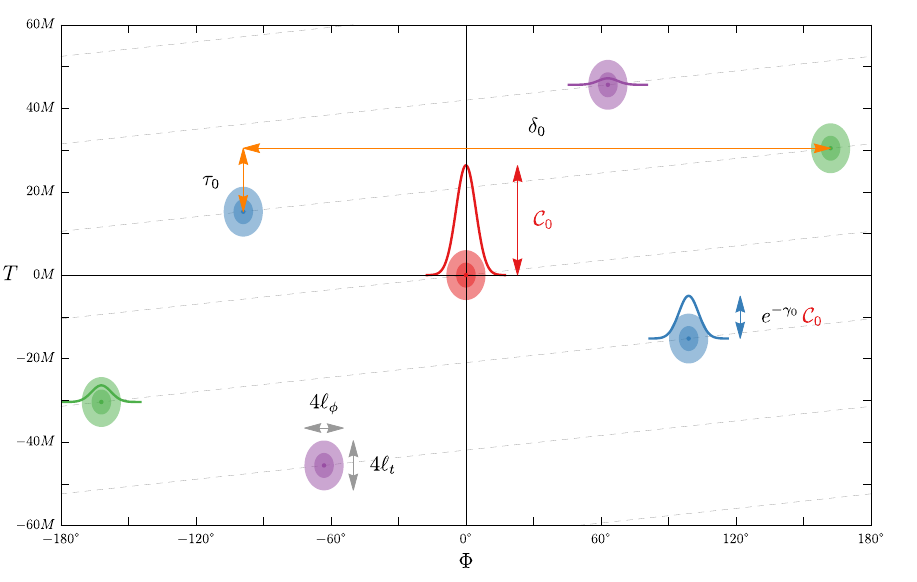}
	\caption{Universal structure in the autocorrelation function $\mathcal{C}(T,\Phi)$ of intensity fluctuations in the photon ring, as a function of the separation in time $T$ and azimuthal angle $\Phi$ around the ring.  This particular plot corresponds to polar observations of random fluctuations in an equatorial disk surrounding a Kerr BH with spin $a/M=94\%$ [Eq.~\eqref{eq:PolarRingCorrelator}]; the same structure holds to leading order in small observer inclination [Eq.~\eqref{eq:Cbar}] and for all spin.  Strong lensing by the BH enables a single source to connect to a given observer along multiple paths (Fig.~\ref{fig:PhotonRingShell}, left), giving rise to correlations within the photon ring (Fig.~\ref{fig:PhotonRingShell}, right).  These correlations display a universal structure governed by the critical exponents $\gamma$, $\delta$, and $\tau$ [Eqs.~\eqref{eq:GeneralCriticalExponents}] that govern BH lensing.  Two light rays that are emitted from the same source and circumnavigate the BH $k$ and $k'$ times, respectively, before reaching the observer contribute to a peak in the autocorrelation $\mathcal{C}(T,\Phi)$ labeled by $m=k-k'$.  Here, we display the $\ab{m}=0,1,2,3$ peaks in red, blue, green and purple, respectively.  The dashed grey lines (with $\Phi=\pm180^\circ$ identified) connect peaks with neighboring values of $m$.  All the peaks share an identical profile that depends on the source statistics.  In particular, the peak width is set by the correlation length of fluctuations in the source; for this plot, we used the correlation lengths $\ell_t$ and $\ell_\phi$ [Eq.~\eqref{eq:Source2PF}] inferred from the GRMHD-simulated accretion flow in Fig.~\ref{fig:GRMHD}.  On the other hand, the locations and relative heights of the peaks are fixed by the BH parameters, with each successive peak an echo of its predecessor, suppressed by $e^{-\gamma}$ and translated by $(\tau,\delta)$ (colored arrows).  Observations of this correlation structure in the photon ring of a BH could therefore provide a measurement of its critical exponents $\tau$ and $\delta$, which would in turn produce estimates of its mass and spin.}
	\label{fig:Autocorrelation}
\end{figure}

In this paper, we explore a new BH observable: \textit{the two-point correlation function} (2PF) \textit{of intensity fluctuations on the photon ring}, $\mathcal{C}=\av{\Delta I(t,\varphi)\,\Delta I(t',\varphi')}$, where $t$ is the observation time and $\varphi$ the angle around the ring.  Intensity fluctuations in the observer sky depend on 1 time and 2 spatial dimensions; we propose, however, to integrate out the direction perpendicular to the critical curve and effectively view the BH as a fluctuating (1+1)-dimensional (unresolved) ring in the sky, on which we study the 2PF.  This correlation function depends on properties of both the BH and its surrounding emission, and exhibits a universal (i.e., matter-independent) structure that is governed by the triplet of critical exponents $\cu{\gamma,\tau,\delta}$.  Previous studies have explored ways to infer BH mass and spin from multiple correlated images of localized sources such as compact infalling or orbiting emitters \cite{Broderick2005,Moriyama2015,Saida2017,Gralla2018,Moriyama2019,Tiede2020,Wong2020}.  Here, we consider the case of stochastic emission from an extended source as a model for emission from the turbulent accretion flow onto a BH \cite{Balbus1991,Yuan2014}.  We analytically compute the 2PF for a toy model of stochastic equatorial emission and show how to separate astrophysical features related to fluctuations in the source from universal features related to the BH (Fig.~\ref{fig:Autocorrelation}).

The rest of this paper is organized as follows.  In Sec.~\ref{sec:Review}, we briefly review some essential properties of null geodesics in the Kerr geometry, with an emphasis on the photon shell and ring.  In Sec.~\ref{sec:IntensityFluctuations}, we present and discuss general properties of the 2PF of intensity fluctuations.  Then, in Sec.~\ref{sec:SourceFluctuations}, we numerically compute the source emissivity 2PF in a full general-relativistic magnetohydrodynamic (GRMHD) simulation and estimate its characteristic correlation length in Boyer-Lindquist radius, azimuthal angle, and time.  Next, in Sec.~\ref{sec:PolarObserver}, we analytically compute the intensity fluctuation 2PF in the case of a polar observer.  We generalize to an observer at arbitrary inclination in Sec.~\ref{sec:GenericInclination}, before expanding to first order in small inclination in Sec.~\ref{sec:SmallInclination}.  Finally, in Sec.~\ref{sec:Observations}, we discuss the prospects for measuring the intensity fluctuation 2PF with the EHT, taking its limitations into account.

\section{Photon ring, photon shell, and critical exponents}
\label{sec:Review}

In this section, we present a brief overview of the basic concepts that we will need for our calculation.  We recently reviewed the photon shell in Ref.~\cite{Johnson2020} and illustrated it in Fig.~2 therein.  Its existence implies that images of sources near a BH will generically feature a bright, narrow ring of light: the photon ring.  In the optically thin limit, this ring is composed of an infinite (in principle) sequence of subrings with a universal (i.e., matter-independent) structure controlled by three critical exponents $\gamma$, $\delta$, and $\tau$, which were recently derived in Refs.~\cite{Johnson2020,GrallaLupsasca2020a}.  After summarizing these facts, we introduce the polar coordinates $(\rho,\varphi)$ on the sky of a distant observer in terms of which our results will be expressed.  Finally, we review how BH images are reconstructed from the electric field 2PF, which is the observable that a radio interferometer directly measures, and discuss higher-point correlations.

\subsection{The Kerr geometry and its photon shell}
\label{sec:PhotonShell}

In Boyer-Lindquist coordinates $(t,r,\theta,\phi)$, the Kerr metric has line element\footnote{We use geometric units $G=c=1$ and assume $0\le a<M$ throughout.}
\begin{gather}
	ds^2=-\frac{\Delta}{\Sigma}\pa{\ed t-a\sin^2{\theta}\ed\phi}^2+\frac{\Sigma}{\Delta}\ed r^2+\Sigma\ed\theta^2+\frac{\sin^2{\theta}}{\Sigma}\br{\pa{r^2+a^2}\ed\phi-a\ed t}^2,\\
	\Delta(r)=r^2-2Mr+a^2,\qquad
	\Sigma(r,\theta)=r^2+a^2\cos^2{\theta}.
\end{gather}
Kerr geodesics admit four independent conserved quantities: the invariant mass $\mu$, the energy $-p_t$ (which is conserved due to stationarity), the azimuthal angular momentum $p_\phi$ (which is conserved due to axisymmetry), and Carter's constant $Q=p_\theta^2-\cos^2{\theta}[a^2(p_t^2-\mu^2)-p_\phi^2\csc^2{\theta}]$.  Geodesic motion in the Kerr spacetime is therefore completely integrable.  Null geodesics ($\mu=0$) are independent of energy, and characterized only by their energy-rescaled azimuthal angular momentum $\lambda=-p_\phi/p_t$ and Carter constant $\eta=Q/p_t^2$.

The Kerr geometry admits a special family of unstable, bound photon orbits \cite{Teo2003}, which span a region of spacetime known as the photon shell \cite{Johnson2020}.  In Boyer-Lindquist coordinates, each of these orbits is at a fixed radius $\tr_-\le\tr\le\tr_+$, where
\begin{align}
	\tr_\pm=2M\br{1+\cos\pa{\frac{2}{3}\arccos\pa{\pm\frac{a}{M}}}}.
\end{align}
The photon shell radius $\tr$ determines both of the conserved quantities for these special geodesics:
\begin{align}
	\label{eq:CriticalGeodesics}
	\tilde{\lambda}=a+\frac{\tr}{a}\pa{\tr-\frac{2\Delta(\tr)}{\tr-M}},\qquad
	\tilde{\eta}=\frac{\tr^3}{a^2}\pa{\frac{4M\Delta(\tr)}{(\tr-M)^2}-\tr}.
\end{align}
Regardless of BH spin, the photon shell always contains spherical bound orbits that can reach the pole.  These have vanishing angular momentum $\tilde{\lambda}(\tr_0)=0$ and orbital radius
\begin{align}
	\label{eq:tr0}
	\tr_0=M+2\sqrt{M^2-\frac{a^2}{3}}\cos\br{\frac{1}{3}\arccos\pa{\frac{1-\frac{a^2}{M^2}}{\pa{1-\frac{a^2}{3M^2}}^{3/2}}}}.
\end{align}
While bound orbits in the photon shell correspond to a measure zero set in the phase space of all geodesics, it is useful to consider geodesics that are nearly bound in the sense that their conserved quantities $\lambda$ and $\eta$ are close to the critical values given in Eq.~\eqref{eq:CriticalGeodesics}.  These near-critical geodesics undergo multiple half-orbits\footnote{An orbit is defined by a full oscillation in the polar angle $\theta$ \cite{Johnson2020}.} within the photon shell and display simple behavior whenever they are close to their associated critical radius $\tr$.  We will use $n$ to denote the number of half-orbits.  In the region $r\approx\tr$, the orbital motion is governed by three critical exponents $\cu{\gamma(\tr),\tau(\tr),\delta(\tr)}$ for each photon shell radius $\tr$ \cite{Johnson2020,GrallaLupsasca2020a}:
\begin{itemize}
	\item Near-critical geodesics spend a long time near their associated critical radius $\tr$.  For a null geodesic that is approaching $\tr$, the ratio of coordinate distances from the critical radius is
	\begin{align}
		\label{eq:RadialConvergence}
		\frac{\dt r_{k+1}}{\dt r_k}\approx e^{-\gamma},
	\end{align}
	where $\dt r_k=r_k-\tr$ denotes the radial deviation of the photon after undergoing $k$ half-orbits.  For a geodesic receding from the critical radius, one needs to flip $k\leftrightarrow (k+1)$ in Eq.~\eqref{eq:RadialConvergence}.
	\item Even as they stay close to their critical radius $\tr$, near-critical geodesics continue to traverse the $t$, $\theta$ and $\phi$ directions.  The time elapsed $\Delta t$ and azimuth swept $\Delta\phi$ per half-orbit approach a constant for high half-orbit number $k$:
	\begin{align}
		\Delta t\approx\tau+\dt t_k,\qquad
		\Delta\phi\approx\dt+\dt\phi_k,
	\end{align}
	with $\dt t_k,\dt\phi_k\sim e^{-k\gamma}\to0$ as $k\to\infty$.
\end{itemize}
The critical exponents were analytically computed in Refs.~\cite{Johnson2020,GrallaLupsasca2020a} and are given by
\begin{subequations}
\label{eq:GeneralCriticalExponents}
\begin{align}
	\gamma&=\frac{4\tr}{a\sqrt{-\tilde{u}_-}}\sqrt{1-\frac{M\Delta(\tr)}{\tr(\tr-M)^2}}K\pa{\frac{\tilde{u}_+}{\tilde{u}_-}},\\
	\delta&=\frac{2}{\sqrt{-\tilde{u}_-}}\br{\frac{\tr+M}{\tr-M}K\pa{\frac{\tilde{u}_+}{\tilde{u}_-}}+\frac{\tilde{\lambda}}{a}\Pi\pa{\tilde{u}_+,\frac{\tilde{u}_+}{\tilde{u}_-}}}+2\pi\Theta(\tr-\tr_0),\\
	\tau&=\frac{2}{a\sqrt{-\tilde{u}_-}}\pa{\tr^2\pa{\frac{\tr+3M}{\tr-M}}K\pa{\frac{\tilde{u}_+}{\tilde{u}_-}}-a^2\tilde{u}_-\br{E\pa{\frac{\tilde{u}_+}{\tilde{u}_-}}-K\pa{\frac{\tilde{u}_+}{\tilde{u}_-}}}},
\end{align}
\end{subequations}
where $K$, $E$, and $\Pi$ respectively denote the complete elliptic integrals of the first, second, and third kinds, while $\Theta$ denotes the Heaviside theta function and $\tilde{u}_\pm$ are the quantities
\begin{align}
	u_\pm=\triangle_\theta\pm\sqrt{\triangle_\theta^2+\frac{\eta}{a^2}},\qquad
	\triangle_\theta=\frac{1}{2}\pa{1-\frac{\eta+\lambda^2}{a^2}},
\end{align}
evaluated on their critical values, obtained by plugging in Eq.~\eqref{eq:CriticalGeodesics}.

\subsection{The observer sky and the photon ring}
\label{sec:RadiativeTransfer}

\begin{figure}[!ht]
	\centering
	\includegraphics[width=\textwidth]{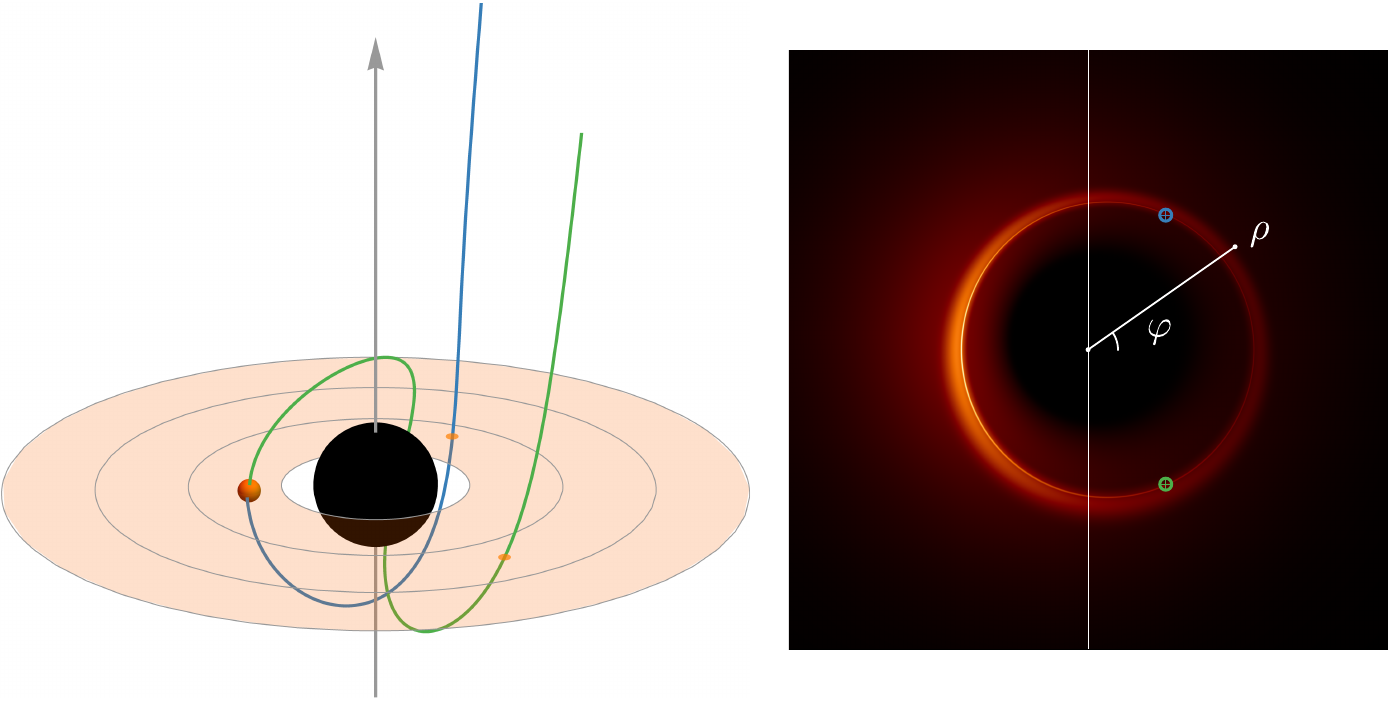}
	\caption{Left: Kerr black hole with spin $a/M=94\%$, surrounded by a geometrically and optically thin equatorial accretion disk terminating at the ISCO radius $r_\mathrm{ms}$.  Right: image of the disk seen by a far observer at an inclination $\theta_o=17^\circ$, assuming a simple stationary and axisymmetric source profile \cite{GrallaLupsasca2020d}.  Light rays that execute multiple orbits around the black hole intersect the emission region multiple times, accruing additional intensity at every crossing (orange dots on the left) according to Eq.~\eqref{eq:RadiativeTransfer}, and resulting in a brightness enhancement near the critical curve: the photon ring.  The polar coordinates \eqref{eq:ScreenCoordinates}, illustrated on the right image, are generically offset from the ring centroid.  Due to the strong lensing in the photon shell, light rays shot back from different times and positions on the photon ring can connect to the same spacetime event in the bulk (blue and green dots on the right image, corresponding to blue and green rays in the bulk).  As a result, black hole images display autocorrelations in the photon ring.}
	\label{fig:PhotonRingShell}
\end{figure}

Bardeen \cite{Bardeen1973} introduced a convenient choice of Cartesian coordinates $(\alpha,\beta)$ on the sky of a distant observer at large radius $r_o\to\infty$ and polar inclination $\theta_o$ from the spin axis of a BH.  Following Ref.~\cite{Johnson2020}, in this paper, we will prefer to use the associated polar coordinates $(\rho,\varphi)$ (Fig.~\ref{fig:PhotonRingShell} right).  A photon reaching the observer with conserved quantities $(\lambda,\eta)$ appears in the sky at position
\begin{align}
	\label{eq:ScreenCoordinates}
	\rho=\frac{1}{r_o}\sqrt{a^2\pa{\cos^2{\theta_o}-u_+u_-}+\lambda^2},\qquad
	\cos{\varphi}=-\frac{\lambda}{r_o\rho\sin{\theta_o}}.
\end{align}
Via this map, the critical curve $\{\tilde{\lambda}(\tr),\tilde{\eta}(\tr)\}$ in the space of photon conserved quantities defines a closed curve $\cu{\tilde{\rho}(\tr),\tilde{\varphi}(\tr)}$ in the observer sky.  This curve separates those light rays that, when shot backwards from the observer sky into the geometry, eventually cross the event horizon (inside the curve) from those that escape to asymptotic null infinity (outside the curve).  Strongly lensed images of emission surrounding the BH, which arise from photons that execute multiple half-orbits $n$, appear exponentially close in $n$ to this critical curve.  While the critical curve is almost a circle at low spin and/or low inclination, the origin of the coordinate system \eqref{eq:ScreenCoordinates} is displaced from its center \cite{Takahashi2004,Johannsen2010,GrallaLupsasca2020c}.

The intensity of light received at a particular time and location in the sky $(t,\rho,\varphi)$ is computed as follows.  In the absence of emission (or absorption), the specific intensity $I_\nu$ varies as the third power of the frequency $\nu$ along a beam of radiation \cite{Lindquist1966}, so $I_\nu/\nu^3$ is conserved along the beam.\footnote{The description of radiation by light rays is valid to leading order in the eikonal approximation.  In the case of EHT observations of M87*, the ratio of the observation wavelength (1.3~mm) to the BH size is $\sim10^{-14}$.}  As such, the observed and emitted specific intensities $I_{\nu_o}$ and $I_{\nu_s}$ are related by radiative transport as
\begin{align}
	\label{eq:RadiativeTransport}
	I_{\nu_o}=g^3I_{\nu_s},
\end{align}
where the redshift $g=\nu_o/\nu_s$, which is the ratio of observed to emitted frequency of the radiation, depends on both the trajectory of the beam and the four-velocity of the emitting matter it intersects.  The emitted specific intensity $I_{\nu_s}$ follows from the radiative transfer equation.  For optically thin sources, absorption effects may be neglected and the invariant radiative transfer equation is \cite{Davis2020}
\begin{align}
	\label{eq:RadiativeTransfer}
	\frac{d}{d\sigma}\pa{\frac{I_\nu}{\nu^3}}=\frac{J_\nu}{\nu^2},
\end{align}
where $\sigma$ is an affine parameter along the null geodesic $x^\mu(\sigma,t,\rho,\varphi)$ ray-traced back from time $t$ and position $(\rho,\varphi)$ in the observer sky, and $J_\nu$ is the source emissivity, which depends only on local properties of the accretion flow.  For simplicity, we take $J_\nu$ to be a scalar, which is akin to assuming isotropic emission at every point.  In general, the emissivity $J_\nu$ varies with frequency, but we will assume throughout this paper that $J\equiv J_\nu$ is frequency-independent, i.e., that the source is broadband with a flat spectrum.\footnote{Because the 230~GHz images of Sgr~A* and M87* have sizes comparable to the expected diameters of their photon rings \cite{Doeleman2008,EHT2019d}, both are likely optically thin at this frequency.  In addition, both sources have a relatively flat spectrum at this frequency \cite{Bower2015,Chael2019,EHT2019f}.} By this, we mean that the emissivity does not vary significantly over roughly an order of magnitude in frequency for both noncritical and near-critical photons.  We have verified this assumption both numerically, using a selection of GRMHD simulations, and analytically in the context of a circularly orbiting equatorial disk, using the methods of Ref.~\cite{Gates2020b}.  Integrating Eq.~\eqref{eq:RadiativeTransfer} over the portion of a light ray that intersects the source results in the specific intensity loaded onto the ray,
\begin{align}
	\label{eq:SourceIntensity}
	I_{\nu_s}=\int J\,\nu_s\ed\sigma
	=\int J\ed s,
\end{align}
where $\ed s$ is the infinitesimal spatial distance traversed by a photon with local frequency $\nu_s$ in affine interval $\ed\sigma$ as measured in the rest frame of the source.  In summary, to compute the observed intensity at a given time and location on the sky, we shoot a null geodesic back into the BH geometry with conserved quantities $(\lambda,\eta)$ determined by the sky position $(\rho,\varphi)$, and whenever the light ray intersects the emission region, we load photons onto it according to the local emissivity $J$, weighting all contributions with a redshift-dependent factor (Fig.~\ref{fig:PhotonRingShell}).

\subsection{Electric field correlation functions and black hole images}

Currently, producing an image $I(t,\rho,\varphi)$ of a BH is only feasible using VLBI.  In this technique, an astronomical source emitting radiation is assumed to be spatially and temporally incoherent.  Despite this intrinsic incoherence, the electric field measured by an observer far from the source exhibits spatial coherence and encodes the source's brightness distribution.  More precisely, a radio interferometer measures the time-averaged cross-spectrum of the two circularly polarized electric fields at different sites, sampling the ``complex visibility'' of the source (e.g., Ref.~\cite{Thompson2017}),
\begin{align}
	\label{eq:ElectricField2PF}
	V_{ij,\nu}=\frac{1}{2}\pa{\av{E_{i,R,\nu}E_{j,R,\nu}^\ast}+\av{E_{i,L,\nu}E_{j,L,\nu}^\ast}},
\end{align}
where $L$ and $R$ refer to the polarization of the feeds, the indices $i$ and $j$ label the elements of the interferometer (i.e., the telescopes in the VLBI array), the asterisk means complex conjugation of the complex electric field, which is sampled at frequency $\nu$, and the angle brackets denote a time average.  Because we focus on broadband emission near black holes, we will suppress the frequency subscript for the remainder of this discussion.

Under very mild assumptions, the van Cittert--Zernike theorem guarantees that this radio visibility is equal to the Fourier transform of a snapshot image of the source in the sky,
\begin{align}
	V(\vec{u})=\int I(\vec{x})e^{-2\pi i\vec{u}\cdot\vec{x}}\ed^2\vec{x},
\end{align}
where $\vec{x}$ is a dimensionless image coordinate measured in radians, while $\vec{u}=\vec{u}_{ij}$ is the dimensionless vector baseline: the distance $\ab{\vec{x}_i-\vec{x}_j}$ between two telescope sites, projected in the plane perpendicular to the line of sight, and measured in units of the observation wavelength $1/\nu$.  The three other choices of cross-correlation functions in Eq.~\eqref{eq:ElectricField2PF} recover the Fourier transforms of the three other Stokes parameters $Q$, $U$, and $V$ describing the polarimetric image of the source.

An immediate question presents itself:  Why not measure higher-point correlation functions of the electric field?  While the first image of M87* occupies only a few kilobytes, the EHT had to record petabytes of data in order to produce it.  In particular, it recorded the local electric field $E_i(\vec{x}_i,t)$ at every telescope site; it is therefore tempting to ask what further information may be encoded in higher-point functions of the electric field, $\langle E_1 E_2\cdots E_n\rangle$ for $n>2$.  In order to answer this question, one must be careful to distinguish between microscopic and macroscopic fluctuations in the source and their associated timescales.

Light rays that cross the source sample many uncorrelated \textit{microscopic} (thermal) fluctuations, which emit band-limited noise with a characteristic timescale of the inverse observing bandwidth ($1/\Delta\nu\sim10^{-9}\,{\rm s}$).  As a result, the complex electric field $E_i(\vec{x}_i,t)$ measured at each site is a sum of many independent emitters and is therefore (by the central limit theorem) a zero-mean Gaussian random field.  Hence, the two-point function \eqref{eq:ElectricField2PF} fully encodes the statistical properties of the source over these microscopic timescales.  In particular, higher-point functions of the electric field are given by Wick's theorem as sums of products of the 2PF (provided that the brackets denote a time average over timescales that are not sufficiently long to probe the macroscopic fluctuations).

In contrast, \textit{macroscopic} fluctuations in the emitting material, such as changes in the plasma properties, produce variations in the image $I(\vec{x})$.  Near a BH and on scales comparable to the photon ring, these fluctuations will have an associated timescale on the order of $M\approx5\pa{\frac{M}{10^6 M_\odot}}\,{\rm s}$.  They also have a nontrivial spatial correlation structure and have coupled spatial and temporal variations that reflect bulk evolution of the emitting material.  The remainder of this paper will focus on these macroscopic fluctuations and the corresponding time-dependent image $I(\vec{x},t)$; accordingly, all time averages will from now on be performed over macroscopic timescales of order $\gtrsim M$.

\section{Two-point function of intensity fluctuations}
\label{sec:IntensityFluctuations}

The main proposal of this paper is to measure and study the 2PF of intensity fluctuations in the observer sky, which is parametrized by time and polar coordinates $(t,\rho,\varphi)$.  The (three-dimensional) 3D correlator is defined in terms of image fluctuations $\Delta I(t,\rho,\varphi)\equiv I(t,\rho,\varphi)-\av{I(t,\rho,\varphi)}$ as
\begin{align}
	\label{eq:IntensityFluctuation2ptFunction}
	\mathcal{C}_\mathrm{3D}(t,t',\varphi,\varphi',\rho,\rho')=\av{\Delta I(t,\rho,\varphi)\,\Delta I(t',\rho',\varphi')}.
\end{align}
This two-point correlator is expected to display an intricate structure that encodes information about the BH parameters, since a single spacetime event in the vicinity of a BH produces multiple (formally, an infinite number of) images in the observer sky that are significantly separated both temporally and spatially.  This phenomenon occurs because any source point in the bulk connects to every observer via an infinite number of different null geodesics that complete an arbitrary number of half-orbits in the photon shell before reaching the observer.

These strongly lensed light rays approach the critical curve $\cu{\tilde{\rho}(\tr),\tilde{\varphi}(\tr)}$ in the sky exponentially fast in the half-orbit number $n$.  The high-order, indirect images can therefore be treated collectively as a single narrow ring.  Operationally, such a treatment is implemented in the correlator by integrating over the radial extent of the photon ring in the observer sky, effectively treating it as an unresolved ring.  Assuming that the fluctuations are stationary (i.e., have time-independent statistics), the resulting photon ring correlator depends only on the autocorrelation time $T=t'-t$:
\begin{align}
	\label{eq:RingCorrelator}
	\mathcal{C}(T,\varphi,\varphi')=\int\rho\ed\rho\int\rho'\ed\rho'\av{\Delta I(t,\rho,\varphi)\,\Delta I(t+T,\rho',\varphi')}.
\end{align}
It is possible to further integrate over the angles around the photon ring, effectively treating the source as pointlike and focusing only on autocorrelations in the time domain:
\begin{align}
\label{eq:1DCorrelator}
	\mathcal{C}_\mathrm{1D}(T)=\int\ed\varphi\int\ed\varphi'\,\mathcal{C}(T,\varphi,\varphi').
\end{align}

\section{Source fluctuations}
\label{sec:SourceFluctuations}

The time-dependent intensity in the observer sky is computed by evaluating Eqs.~\eqref{eq:RadiativeTransport} and \eqref{eq:SourceIntensity} for a given source distribution.  In order to compute $\mathcal{C}(T,\varphi,\varphi')$, therefore, we need information about the source statistics.  In Sec.~\ref{sec:PolarObserver}, we will consider a simple, analytically tractable toy model for random equatorial emission, with a specific (Gaussian) profile for the emissivity 2-point correlation function in the $t$, $r$, and $\phi$ directions.  In this section, we guide our choice of parameters for this emissivity 2PF by considering a GRMHD BH accretion flow simulation.  We are particularly interested in the correlation lengths in the different spacetime directions on which the source profile depends.  In contrast, previous studies of correlations in GRMHD accretion flows have focused on density, temperature, and magnetic field (rather than emissivity), especially in the context of convergence studies \cite{Guan2011,Shiokawa2012}.\footnote{In general, emissivity depends on frequency, so redshift effects (due to the gravitational field of the BH and Doppler beaming from the motion of the fluid) may be significant.  In our simulation, however, we find that the redshift factors between the local frame of emission and the observer are of order two, and moreover, that correlation statistics do not vary significantly over the corresponding range of frequencies near the 230~GHz observing frequency.  As a result, we may neglect the frequency dependence of the emissivity 2PF, as described in Sec.~\ref{sec:RadiativeTransfer}.}

The simulation was produced with the \texttt{iharm3d} code \cite{Gammie2003} and was initialized to correspond to a magnetically arrested disk accretion flow \cite{Igumenshchev2003,Narayan2003} around a BH of spin $a/M=94\%$.  The synchrotron emissivity at 230~GHz was computed according to Ref.~\cite{Leung2011} assuming a thermal electron distribution function.  Electron temperatures were computed from the bulk fluid internal energy according to the prescription described in Ref.~\cite{EHT2019e} with $r_\mathrm{high}=40$.  We scale the mass and length units of the GRMHD simulation to target an M87-like observation with compact 230~GHz flux $\approx0.7\,{\rm Jy}$, as in Ref.~\cite{EHT2019e}.  We evaluate the emissivity $J$ in the midplane of the simulation on snapshots spaced at time intervals of $0.5M$.  In addition to depending on the plasma number density and electron temperature, $J$ is also a function of the angles between the line of sight and both the magnetic field (pitch angle) as well as the fluid velocity (fluid-frame frequency)---hence, it is a tensorial quantity.  Since we focus on the low-inclination regime and assume that all statistical properties of the accretion flow inherit the axisymmetry of the underlying Kerr geometry, we will neglect corrections due to $\phi$-dependence of the emissivity on the pitch angle and the fluid-frame frequency.

\begin{figure}[!ht]
	\centering
	\includegraphics[width=\textwidth]{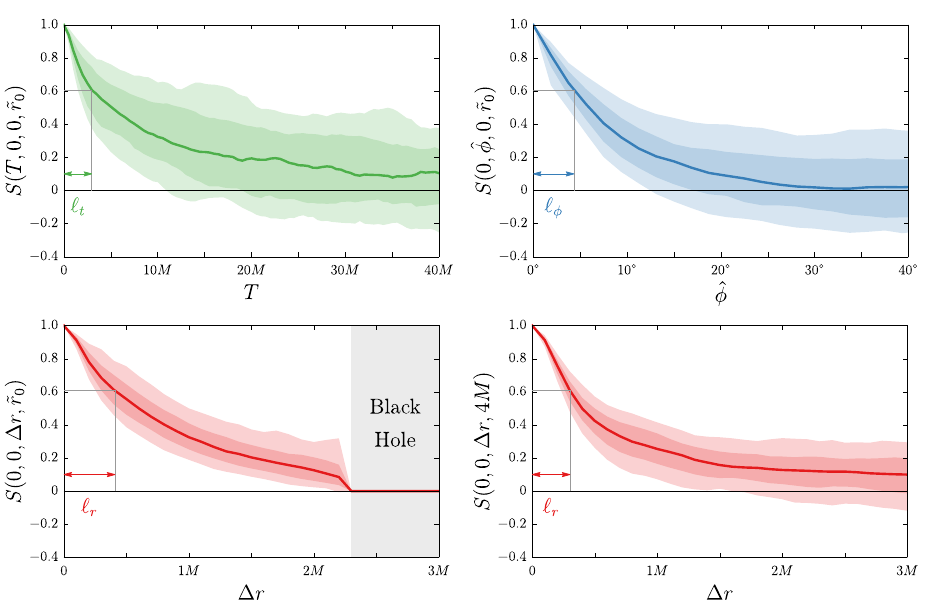}
	\caption{Normalized correlation function $S$ of fluctuations in 230~GHz synchrotron emissivity computed in the midplane of a numerical GRMHD simulation.  The configuration consists of a magnetically arrested disk accreting onto a Kerr BH of spin $a/M\approx94\%$.  Top left: correlation function in $T$ assuming $\hat{\phi}=\Delta r=0$.  Top right: azimuthal correlation in $\hat{\phi}$ for $T=\Delta r=0$.  Both panels in the top row are evaluated at $\bar{r}=\tr_0\approx2.5M$.  Bottom row: radial correlation in $\Delta r$ evaluated at two different radii $\bar{r}=\tr_0$ and $\bar{r}=4M$.  The emissivity is set to zero behind the event horizon $r=r_+$, so correlations vanish for $\Delta r\ge2(\bar{r}-r_+)$. This occurs in the shaded region in the bottom left panel. For a unit-height Gaussian, the standard deviation width (1$\sigma$) is achieved at height $e^{-1/2}\approx60.65\%$. The correlation lengths at $\bar{r}=\tr_0$ are $\ell_t\approx3.0M$, $\ell_\phi\approx4.3^\circ$, and $\ell_r\approx0.4M$ (while $\ell_r\approx0.3M$ at $\bar{r}=4M$).
	}
	\label{fig:GRMHD}
\end{figure}

We study the emissivity 2PF as a function of four variables:
\begin{align}
	T=t'-t,\qquad
	\hat{\phi}=\phi'-\phi,\qquad
	\Delta r=r'-r,\qquad
	\bar{r}=\frac{r+r'}{2}.
\end{align} 
In simulations, the emissivity is often a sharp function of radius, and so the properties of the correlation function may depend on $\bar{r}$.  We compare the statistics of the $T$, $\hat{\phi}$, and $\Delta r$ correlations for different values of $\bar{r}$.  We effectively assume separability\footnote{We assume that $S(T,\hat{\phi},\Delta r,\bar{r})$ is separable in Boyer-Lindquist coordinates in order to obtain crude estimates for the correlation length in emissivity fluctuations.  We expect a more refined analysis to show that this simplistic assumption breaks down because of various effects; for instance, the flow velocity defines a special radius-dependent direction in the $(t,\phi)$ plane.  We consider potential observational consequences of this assumption in Sec.~\ref{sec:Observations}.} by computing the correlations in $T$ and $\hat{\phi}$ at fixed radius (so that $r=r'$).

We evaluate the synchrotron emissivity in the midplane $\bar{J}(t,r,\phi)\equiv\bar{J}(t,r,\phi;\theta=\pi/2)$ directly from the numerical simulation.  Note that $\bar{J}(t,r,\phi)=\bar{J}(t,r,\phi+2\pi)$ by the azimuthal periodicity of the domain.  The normalized emissivity 2PF is given by
\begin{align}
	S(T,\hat{\phi},\Delta r,\bar{r})=\frac{\av{\Delta \bar{J}(t+T,\phi+\hat{\phi},\bar{r}+\Delta r/2)\,\Delta \bar{J}(t,\phi,\bar{r}-\Delta r/2)}}{\av{\br{\Delta \bar{J}(t,\phi,\bar{r})}^2}},
\end{align}
where the mean-subtracted emissivity $\Delta\bar{J}\equiv \bar{J}-\av{\bar{J}}_{t,\phi}$ encodes the fluctuations about $\av{\bar{J}}_{t,\phi}$, the time-and-azimuthal-average of the emissivity in the simulation.

In Fig.~\ref{fig:GRMHD}, we plot $S(T,0,0,\bar{r})$ as a function of $T$, $S(0,\hat{\phi},0,\bar{r})$ as a function of $\hat{\phi}$, and $S(0,0,\Delta r,\bar{r})$ as a function of $\Delta r$, for various fixed values of $\bar{r}$.  We define the correlation length as the width away from the maximum at a height of $e^{-1/2}$, which for a Gaussian would correspond to 1 standard deviation $\sigma$.  At $\bar{r}=\tr_0$, we extract from our simulation the correlation lengths $\ell_t\approx3.0M$, $\ell_\phi\approx4.3^\circ$, and $\ell_r\approx0.4M$ (Fig.~\ref{fig:GRMHD} also displays the correlation length $\ell_r\approx0.3M$ found at $\bar{r}=4M$).

Note that, although we compute its correlation function in the midplane, the emissivity is a volumetric quantity with support in the geometrically thick disk.  In the remainder of this paper, we will consider an equatorial thin disk model with volumetric emissivity $J(t,r,\phi)\delta(\theta-\pi/2)$, corresponding to an effective surface emissivity $J(t,r,\phi)$ that is designed to reproduce the observable statistics of the thick disk model.  We will assume that the correlation structure of $\bar{J}$ (the emissivity in the midplane of a geometrically thick disk) is a reasonable proxy for correlations in $J$. A finite thickness would introduce partial cancellation of fluctuations that would reduce the 2PF for all separations, including $\mathcal{C}(T=0, \varphi=\varphi')$.

\section{Toy emission model: polar observer}
\label{sec:PolarObserver}

In this section, we focus on the special case of a polar observer, in which the large-$n$ contribution to the autocorrelation \eqref{eq:RingCorrelator} (from light rays that circumnavigate the BH multiple times) is especially simple.  Since this configuration preserves axisymmetry, the correlator depends only on $\Phi=\varphi'-\varphi$:
\begin{align}
	\label{eq:PolarRingCorrelator}
	\mathcal{C}(T,\Phi)=\int\rho\ed\rho\int\rho'\ed\rho'\av{\Delta I(t,\rho,\varphi)\,\Delta I(t+T,\rho',\varphi+\Phi)}.
\end{align}
We consider an optically thin annular emission region localized in the equatorial plane $\theta=\pi/2$ with local emissivity $J(t,r,\phi)\delta\pa{\theta-\frac{\pi}{2}}$ satisfying
\begin{align}
	\label{eq:Source2PF}
	\av{\Delta J(t,r,\phi)\,\Delta J(t',r',\phi')}=
	\begin{cases}
		\mathcal{J}G_{\ell_t}(t-t')G_{\ell_r}(r-r')G_{\ell_\phi}^\circ(\phi-\phi')
		&\text{if }r_\mathrm{min}\le r,r'\le r_\mathrm{max},\\ 0&\text{otherwise}.
	\end{cases}
\end{align}
Here, $r_\mathrm{max}-r_\mathrm{min}=W$ is the width of the annular equatorial disk, $\Delta J(t,r,\phi)\equiv J(t,r,\phi)-\av{J(t,r,\phi)}$ is the source fluctuation, and we introduced the distributions
\begin{align}
	G_\ell(z)=\frac{1}{\sqrt{2\pi\ell^2}}e^{-\frac{z^2}{2\ell^2}},\qquad
	G_\ell^\circ(\phi)=\frac{1}{2\pi I_0\pa{1/\ell^2}e^{-1/\ell^2}}e^{-\frac{2}{\ell^2}\sin^2\pa{\frac{\phi}{2}}}.
\end{align}
These are, respectively, the Gaussian distribution and its analogue for a periodic variable, known as the von Mises distribution,\footnote{Whenever $0<\epsilon=\ab{\phi-\phi'}\mod2\pi\ll1$, the von Mises and Gaussian distributions agree to leading order in $\epsilon$.} with normalization chosen to ensure that
\begin{align}
	\int_{-\infty}^\infty G_\ell(z)\ed z=\int_0^{2\pi}G_\ell^\circ(\phi)\ed\phi
	=1.
\end{align}
The parameters $\ell_t$, $\ell_r$, and $\ell_\phi$ are the correlation lengths of the fluctuating source in the $t$, $r$, and $\phi$ directions, respectively.  Our goal is to analytically compute the contribution to the correlator \eqref{eq:PolarRingCorrelator} from photons undergoing multiple half-orbits around the BH.  Such near-critical geodesics are close to the photon shell and display the universal properties discussed in Sec.~\ref{sec:PhotonShell}, with a significant simplification due to setting $\theta_o=0$.  In this special case, only photons with zero angular momentum, which execute multiple orbits near the radius $\tr_0$ in the photon shell, may reach the observer.  We assume here the emission region contains the photon shell, so that $r_\mathrm{min}<\tr_0<r_\mathrm{max}$.  In the observer sky, the near-critical regime is defined by $\rho=\tilde{\rho}_0+\dt\rho$ with $\dt\rho/\tilde{\rho}_0\ll1$, where
\begin{align}
	\label{eq:CriticalCurveRadius}
	\tilde{\rho}_0=\sqrt{\frac{\tr_0^3}{a^2}\br{\frac{4M\Delta(\tr_0)}{\pa{\tr_0-M}^2}-\tr_0}+a^2}
\end{align}
is the radius of the perfectly circular critical curve in the sky.  Minding a subtlety in $\delta$ [Eq.~(66) of Ref.~\cite{GrallaLupsasca2020a}], the critical exponents corresponding to the single observable photon shell radius $\tr_0$ are
\begin{subequations}
\label{eq:PolarCriticalExponents}
\begin{align}
	\gamma_0&=\frac{4\tr_0}{\sqrt{\tilde{\rho}_0^2-a^2}}\sqrt{1-\frac{M\Delta(\tr_0)}{\tr_0(\tr_0-M)^2}}K\pa{\frac{a^2}{a^2-\tilde{\rho}_0^2}},\\
	\delta_0&=\pi+\frac{2a}{\sqrt{\tilde{\rho}_0^2-a^2}}\pa{\frac{\tr_0+M}{\tr_0-M}}K\pa{\frac{a^2}{a^2-\tilde{\rho}_0^2}},\\
	\tau_0&=\frac{2}{\sqrt{\tilde{\rho}_0^2-a^2}}\cu{\tr_0^2\pa{\frac{\tr_0+3M}{\tr_0-M}}K\pa{\frac{a^2}{a^2-\tilde{\rho}_0^2}}-2a^2\br{E\pa{\frac{a^2}{a^2-\tilde{\rho}_0^2}}-K\pa{\frac{a^2}{a^2-\tilde{\rho}_0^2}}}}.
\end{align}
\end{subequations}

The observed specific intensity of the equatorial disk is obtained via the procedure outlined in Sec.~\ref{sec:RadiativeTransfer}.  The radiative transport equation \eqref{eq:RadiativeTransport} implies that a light ray shot back from position $(\rho,\varphi)$ in the observer sky collects photons only when it crosses the disk, so
\begin{align}
	 \label{eq:SumOverCrossings}
	I_{\nu_o}(t,\rho,\varphi)=\sum_{k=0}^ng^3I_{\nu_s}\pa{r_s^{(k)}(\rho,\varphi)},
\end{align}
where the sum is taken over the $n$ equatorial crossings of the light ray, which intersects the disk at radii $r_s^{(k)}(\rho,\varphi)$ for $k\in\cu{0,1,\ldots,n}$ (see App.~\ref{app:Lensing} for an exact formula).  The emitted specific intensity at each crossing is computed using the radiative transfer equation \eqref{eq:RadiativeTransfer} as follows.  Let $\Theta$ denote the emission angle of the light ray relative to the local zenith ($\theta$-direction) in the rest frame of the source.  For a light ray crossing the equatorial plane perpendicularly ($\Theta=0$), we have $\ed s=\sqrt{g_{\theta\theta}}\ed\theta=\sqrt{\Sigma}\ed\theta=\sqrt{r^2+a^2\cos^2{\theta}}|_{\theta=\pi/2}\ed\theta=r\ed\theta$.  More generally, if it crosses at an angle $\Theta>0$ from the zenith in the rest frame, then
\begin{align}
	\ed s=\frac{r}{\cos{\Theta}}\ed\theta.
\end{align}
Evaluating Eq.~\eqref{eq:RadiativeTransfer} in the rest frame of the source, where $J_\nu=J(t,r,\phi)\delta\pa{\theta-\frac{\pi}{2}}$, it follows that\footnote{In the covariant formalism for radiative transfer, $I_{\nu_s}(r)=\nu_s\int J(d\sigma/d\theta)\ed\theta$, where $|d\theta/d\sigma|=|p^\theta|=\nu_o\sqrt{\eta}/r^2$ [e.g., by Eq.~(6b) of Ref.~\cite{GrallaLupsasca2020b} evaluated at $\theta=\pi/2$ with $E=\nu_o$].  Hence, $I_{\nu_s}(r)=J(t,r,\phi)r^2/(g\sqrt{\eta})$, which agrees with Eq.~\eqref{eq:CrossingIntensity} in light of Eq.~\eqref{eq:DirectionCosine}.}
\begin{align}
	 \label{eq:CrossingIntensity}
	 I_{\nu_s}(r)=\int J(t,r,\phi)\delta\pa{\theta-\frac{\pi}{2}}\frac{ds}{d\theta}\ed\theta
	 =\frac{r}{\cos{\Theta}}J(t,r,\phi).
\end{align}
Since the effective surface emissivity $J$ is by assumption independent of the emission frequency $\nu_s$ in the rest frame of the source, we will from now on omit the frequency subscripts.  In general, the precise forms of the redshift $g$ and direction cosine $\cos{\Theta}$ depend on the relative velocity of the emitting matter in the disk.  In the model presented here, we will assume that this velocity field is equatorial and axisymmetric.  For example, in the particular case of emitters on corotating circular equatorial orbits, one obtains the simple expression \cite{Cunningham1975,Gralla2018}
\begin{align}
	\label{eq:DirectionCosine}
	\cos\Theta=\pm\frac{g\sqrt{\eta}}{r},
\end{align}
where the upper or lower sign is chosen according to whether the light ray with conserved quantities $(\lambda,\eta)$ is emitted upwards or downwards from the disk and the redshift factor $g(r,\lambda)$, whose form depends on the stability of the orbit, is given in App.~\ref{app:Redshift}.  Together, Eqs.~\eqref{eq:SumOverCrossings} and \eqref{eq:CrossingIntensity} imply that
\begin{align}
	I(t,\rho,\varphi)=\sum_{k=0}^ng^3\frac{r_s^{(k)}}{\cos{\Theta}}J\pa{t_s^{(k)},r_s^{(k)},\phi_s^{(k)}},
\end{align}
where each term in the sum is to be evaluated at the spacetime coordinates of the corresponding equatorial crossing. Note that for circularly orbiting emitters, the redshift factor in $\cos{\Theta}$ [Eq.~\eqref{eq:DirectionCosine}] effectively reduces the power of $g^3$ to $g^2$.  In particular, for a near-critical light ray reaching the polar observer, we have $g\approx \tilde{g}_0=g(\tr_0,\lambda=0)$ and $\cos{\Theta}\approx\cos{\tilde{\Theta}_0=\cos{\Theta}(\tr_0,\lambda=0)}$,\footnote{For circularly orbiting emitters, $\cos{\tilde{\Theta}_0}=\pm\tilde{g}_0\sqrt{\tilde{\eta}(\tr_0)}/\tr_0$.} and thus
\begin{align}
	 \label{eq:NearCriticalIntensity}
	I(t,\tilde{\rho}_0+\dt\rho,\varphi)\approx\tilde{g}_0^3\sum_{k=0}^n\frac{\tr_0}{\cos{\tilde{\Theta}_0}}J\pa{t-\Delta t_0-k\tau_0-\dt t_k,\tr_0+\dt r_k,\varphi-\Delta\phi_0-k\delta_0-\dt\phi_k},
\end{align}
where $n(\dt\rho)$ is the number of half-orbits executed, which obeys\footnote{The subleading ($\dt\rho$-independent) correction to this relation was computed analytically in Ref.~\cite{GrallaLupsasca2020a}.} $n\sim-\ln{\dt\rho/\tilde{\rho}_0}$.  The deviations $\dt t_k$, $\dt r_k$, and $\dt\phi_k$ all vanish as $k\to\infty$; for large $1\ll k<n$, they give small corrections that account for the fact that the geodesic is slightly near-critical (off the photon shell).  The azimuthal winding $\Delta\phi_0=\mathrm{const.}+\O{\dt\rho}$ is a $\varphi$-independent angle accumulated between the observer and the first equatorial crossing (when ray-tracing backwards from the observer sky) that is spin-dependent but irrelevant for the photon ring contribution to the two-point function, as it is also approximately the same for all near-critical photons.  Similarly, the time lapse $\Delta t_0=\mathrm{const.}+\O{\dt\rho}$ is the time elapsed along a photon trajectory from the last crossing of the equatorial plane to the observer (evolving the ray forward in time), which is approximately the same for all near-critical photons.

The intensity fluctuations in the photon ring are obtained by subtracting from both sides of Eq.~\eqref{eq:NearCriticalIntensity} their average, resulting in
\begin{align}
	 \label{eq:IntensityFluctuations}
	\Delta I(t,\tilde{\rho}_0+\dt\rho,\varphi)\approx\tilde{g}_0^3\sum_{k=0}^n\frac{\tr_0}{\cos{\tilde{\Theta}_0}}\Delta J\pa{t-\Delta t_0-k\tau_0-\dt t_k,\tr_0+\dt r_k,\varphi-\Delta\phi_0-k\delta_0-\dt\phi_k},
\end{align}
where $\Delta J$ is the fluctuation of the effective surface emissivity of the disk.  Now we can plug Eq.~\eqref{eq:IntensityFluctuations} into Eq.~\eqref{eq:PolarRingCorrelator}, and use Eq.~\eqref{eq:Source2PF} to obtain the following formula for the photon ring autocorrelations:
\begin{align}
	\mathcal{C}(T,\Phi)&\approx\frac{\mathcal{J}\tr_0^2\tilde{g}_0^6}{\cos^2{\tilde{\Theta}_0}}\int\rho\ed(\dt\rho)\int\rho'\ed(\dt\rho')\sum_{k=0}^{n(\dt\rho)}\sum_{k'=0}^{n'(\dt\rho')}G_{\ell_r}\pa{\dt r_{k'}-\dt r_k}\\
	&\quad\times G_{\ell_t}\br{T-\tau_0(k'-k)-\dt t_{k'}+\dt t_k}G_{\ell_\phi}^\circ\br{\Phi-\delta_0(k'-k)-\dt \phi_{k'}+\dt\phi_k}.\notag
\end{align}
One may now interchange the order of integration and summation, thereby replacing the sum over crossings with a sum over subrings.  Recalling that $n(\dt\rho)\to\infty$ as $\dt\rho\to0$, this results in
\begin{align}
	\label{eq:ToyModelIntermediateStep1}
	\mathcal{C}(T,\Phi)&\approx\frac{\mathcal{J}\tr_0^2\tilde{g}_0^6}{\cos^2{\tilde{\Theta}_0}}\sum_{k=0}^\infty\sum_{k'=0}^\infty\int_{\mathrm{s.r.}k}\rho\ed(\dt\rho)\int_{\mathrm{s.r.}k'}\rho\ed(\dt\rho')\,G_{\ell_r}\pa{\dt r_{k'}-\dt r_k}\\
	&\quad\times G_{\ell_t}\br{T-\tau_0(k'-k)-\dt t_{k'}+\dt t_k}G_{\ell_\phi}^\circ\br{\Phi-\delta_0(k'-k)-\dt \phi_{k'}+\dt\phi_k},\notag
\end{align}
where the limits of integration in the integral $\int_{\mathrm{s.r.}k}$ are the boundaries of the $k^\text{th}$ subring.  The integrations over $\dt\rho$ and $\dt\rho'$ may now be evaluated as follows.  First, we will make the approximation
\begin{align}
	 \label{eq:NegligibleDeviations}
	\dt t_k-\dt t_{k'}\approx\dt\phi_k-\dt\phi_{k'}
	\approx 0,
\end{align}
which holds for large enough $\hat{k}<n$, where $\hat{k}=k$ if $\ab{k'-n/2}<\ab{k-n/2}$, and $\hat{k}=k'$ otherwise; more precisely, due to the exponential falloff of $\dt t_k$, $\dt\phi_k$, Eq.~\eqref{eq:NegligibleDeviations} is valid when
\begin{align}
	\begin{cases}
		Me^{-\hat{k}\gamma_0}\ll\ell_t\text{ and }
		e^{-\hat{k}\gamma_0}\ll\ell_\phi
		&\text{for }1\ll\hat{k}\le n/2,\\
		Me^{-(n-\hat{k})\gamma_0}\ll\ell_t\text{ and }
		e^{-(n-\hat{k})\gamma_0}\ll\ell_\phi
		&\text{for }n/2\le\hat{k}\le n.
	\end{cases}
\end{align}
In contrast, to leading order as $\hat{k}\to\infty$, the argument of the radial Gaussian function vanishes identically; hence, we must must keep track of the next order in $\dt r_k$, as described in detail below.

Finally, we can also approximate the measure near the photon shell as $\rho\ed(\dt\rho)\approx\tilde{\rho}_0\ed(\dt\rho)$.  Under these approximations, the only $\dt\rho$-dependence of the integrand comes in via $\dt r_{k/k'}$, so we only need to compute the contribution
\begin{align}
	\label{eq:Akk'}
	A_{k,k'}=\int_{\mathrm{s.r.}k}\ed(\dt\rho)\int_{\mathrm{s.r.}k'}\ed(\dt\rho')\,G_{\ell_r}\pa{\dt r_{k'}-\dt r_k}.
\end{align}
This factor determines the amplitude of the contribution of the $\{k,k'\}$ summand in (\ref{eq:ToyModelIntermediateStep1}). We prove in App.~\ref{app:Lensing} that
\begin{align}
	\label{eq:RingDemagnification}
	\ed(\dt\rho)=\iota_0e^{-k\gamma_0}\ed(\dt r_k),
\end{align}
where
\begin{align}
	\iota_0=\frac{32\tr_0^4\tilde{\chi}^2e^{-\frac{1}{2} \gamma_0}}{\tilde{\rho}_0\pa{1+\sqrt{\tilde{\chi}}}\Delta(\tr_0)},\qquad
	\tilde{\chi}=1-\frac{M\Delta(\tr_0)}{\tr_0\pa{\tr_0-M}^2}.
\end{align}
The radial integral \eqref{eq:Akk'} that we wish to compute is therefore of the form
\begin{align}
	I_C=\int_{-W/2}^{W/2}\ed x\int_{-W/2}^{W/2}\ed y\,G_{\ell}\pa{x-y+C},
\end{align}
with $C=0$ (but we will use the generalization to nonzero $C$ later).  Assuming $\ell<W$, this integral evaluates to
\begin{align}
	\label{eq:RadialIntegral}
	I_C=W\Lambda\pa{\frac{C}{W}},
\end{align}
where
\begin{align}
	\label{eq:LambdaFunction}
	\Lambda(z)&=\frac{1+z}{2}\erf\pa{\frac{1+z}{\sqrt{2}\ell/W}}-\frac{z}{2}\erf\pa{\frac{z}{\sqrt{2}\ell/W}}+\pa{\frac{\ell}{W}}^2\pa{e^{-\frac{z+1/2}{(\ell/W)^2}}-1}G_{\ell/W}(z)\notag\\
	&\quad+(z\to-z),
\end{align}
and $\erf(z)=(2/\sqrt{\pi})\int_0^ze^{-t^2}\ed t$ denotes the standard error function.  Note that $\Lambda(z)$ achieves its peak value (which is less than unity) at the origin and falls off monotonically in $\ab{z}$, tending to zero as $\ab{z}\to\infty$.  In particular,
\begin{align}
	\Lambda(0)=\erf\pa{\frac{W}{\sqrt{2}\ell}}-\sqrt{\frac{2}{\pi}}\pa{\frac{\ell}{W}}\br{1-e^{-(W/\ell)^2/2}}.
\end{align}
Also, note that for $\ell\ll W$,
\begin{align}
	\Lambda(z)\approx\mathrm{max}\pa{1-\ab{z},0}.
\end{align}
Using these identities, Eq.~\eqref{eq:Akk'} evaluates to
\begin{align}
	A_{k,k'}=\iota_0^2e^{-\gamma_0(k+k')}W\Lambda(0).
\end{align}
Now we can change the order of summation, defining $m=k-k'$ and $2s=k+k'-|m|$ to obtain
\begin{align}
	\mathcal{C}(T,\Phi)&=\pa{\frac{\iota_0\tilde{\rho}_0\tr_0\tilde{g}_0^3}{\cos{\tilde{\Theta}_0}}}^2\mathcal{J}W\Lambda(0)\sum_{m=-\infty}^\infty \sum_{s=0}^\infty e^{-\gamma_0(|m|+2s)}G_{\ell_t}(T+m\tau_0)G_{\ell_\phi}^\circ(\Phi+m\delta_0)\\
	\label{eq:ToyModelIntermediateStep2}
	&=\pa{\frac{\iota_0\tilde{\rho}_0\tr_0 \tilde{g}_0^3}{\cos{\tilde{\Theta}_0}}}^2\frac{\mathcal{J}W\Lambda(0)}{1-e^{-2\gamma_0}}\sum_{m=-\infty}^\infty e^{-\gamma_0|m|}G_{\ell_t}(T+m\tau_0)G_{\ell_\phi}^\circ(\Phi+m\delta_0).
\end{align}
This expression describes a train of peaks in the autocorrelation plane $(T,\Phi)$, localized around  $(m\tau_0,m\delta_0)$ for integer $m$, and with exponentially decaying amplitude as $|m|$ grows.  It is clear that the function $\mathcal{C}(T,\Phi)$ enjoys a discrete symmetry, or self-similarity: it is invariant under
\begin{align}
	T\to T+\tau_0,\qquad
	\Phi\to\Phi+\dt_0,\qquad
	\mathcal{C}\to e^{\pm\gamma_0}\mathcal{C},\qquad
	\pm=\sign(m).
\end{align}
The correlator has both universal and nonuniversal features, but the simple form of the result in Eq.~\eqref{eq:ToyModelIntermediateStep2} allows us to separately extract them from $\mathcal{C}(T,\Phi)$: the location of the peaks in the $(T,\Phi)$ autocorrelation plane, and the ratios between their amplitudes, are determined by universal features of the Kerr geometry: the critical exponents $\gamma_0$, $\delta_0$, and $\tau_0$.  On the other hand, the shape of a single peak, and its width in particular, are determined by the astrophysical details of the accretion flow.  For example, we could have chosen to model the source 2PF in Eq.~\eqref{eq:Source2PF} by Lorentzians instead of Gaussians.  Each peak would have then changed its shape but not its location in the $(T,\Phi)$ plane.  We illustrate the universal self-similar structure of $\mathcal{C}(T,\Phi)$ in Fig.~\ref{fig:Autocorrelation}.

The formula \eqref{eq:ToyModelIntermediateStep2} and its generalization to inclined observers obtained below present promising prospects for observation (see Sec.~\ref{sec:Observations} for further discussion).  Upon measuring $\mathcal{C}(T,\Phi)$, it would be extremely interesting to see whether more than one peak can be observed.  In fact, observing two or more clearly separated peaks would already provide strong evidence of the photon ring: strongly lensed photons that execute multiple orbits around the BH.  Moreover, these peaks must be arranged in the autocorrelation plane in a way that respects the self-similarity described above, a fact that could provide a highly nontrivial test of strong-field General Relativity.  Most interestingly, a measurement of the locations of these peaks provides a novel method to measure BH parameters: \textit{both mass and spin}.  This would be done as follows: the peaks' locations provide information on $\tau_0$ and $\delta_0$.  The critical exponent $\tau_0$ depends weakly on spin \cite{GrallaLupsasca2020a} and thus is a good measure of the BH mass.\footnote{Of course, there are already quite good independent mass estimates for certain BHs such as M87*.}  On the other hand, $\delta_0$ does depend strongly on spin and could thus be used to measure it. 

Finally, we can also integrate over the angular dependence $\Phi$ to isolate the time autocorrelations:
\begin{align}
	\mathcal{C}_\mathrm{1D}(T)&=\int\ed\varphi\int\ed\varphi'~\mathcal{C}(T,\Phi)
	=2\pi\ell_\varphi\pa{\frac{\iota_0\tilde{\rho}_0\tr_0\tilde{g}_0^3}{\cos{\tilde{\Theta}_0}}}^2\frac{\mathcal{J}W\Lambda(0)}{1-e^{-2\gamma_0}}\sum_{m=-\infty}^\infty e^{-\gamma_0|m|}G_{\ell_t}(T+m\tau_0).
\end{align}
This observable has the advantage of being measurable even if the ring's diameter is not resolved.

\begin{figure}[!ht]
	\centering
	\includegraphics[width=\textwidth]{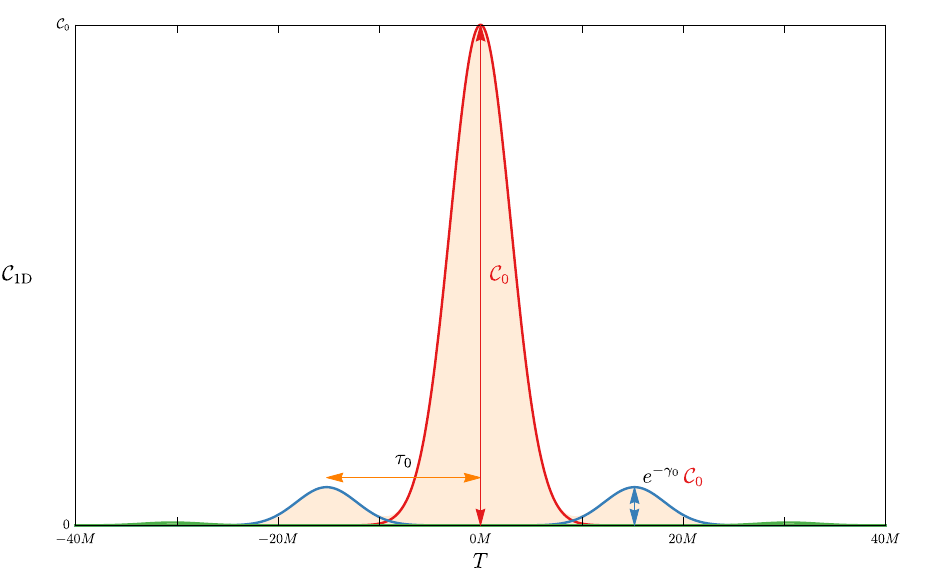}
	\caption{Universal structure in the time autocorrelation function $\mathcal{C}_\mathrm{1D}(T)$ [Eq.~\eqref{eq:1DCorrelator}], with $T$ the time separation, for polar observations of random fluctuations in an equatorial disk surrounding a Kerr BH with spin $a/M=94\%$.  Universal aspects of $\mathcal{C}_\mathrm{1D}(T)$ are governed by the critical exponents $\gamma$ and $\tau$ [Eqs.~\eqref{eq:GeneralCriticalExponents}].  The autocorrelation function consists of a sum (displayed with orange shading) of self-similar peaks localized around $T=m\tau_0$ for every integer $m$, differing only by an overall demagnification factor $e^{-|m|\gamma_0}$.  Here, we show the $|m|=0,1,2$ peaks in red, blue, and green respectively. The profile of each peak, which is nonuniversal, depends on statistical properties of the flow, and is taken here to correspond to our toy model \eqref{eq:Source2PF}.  This structure may be obtained by integrating over $\Phi$ in Fig.~\ref{fig:Autocorrelation}.}
	\label{fig:TimeAutocorrelation}
\end{figure}

\section{Generalization to inclined observer}
\label{sec:GenericInclination}

For an observer at nonzero inclination $\theta_o>0$, the axisymmetry in the observer sky is broken but time-translation symmetry is preserved at the statistical level.  The correlation function will therefore have the general form 
\begin{align}
	\label{eq:InclinedC}
	\mathcal{C}(T,\varphi,\varphi')=\int\rho\ed\pa{\dt\rho}\int\rho'\ed\pa{\dt\rho'}\av{\Delta I(t,\tilde{\rho}+\dt\rho,\varphi)\,\Delta I(t+T,\tilde{\rho}+\dt\rho',\varphi')}.
\end{align}

Generalizing the analysis of Sec.~\ref{sec:PolarObserver}, we can write the intensity fluctuation as
\begin{align}
	\label{eq:InclinedIntensityFluctuations}
	\Delta I(t,\tilde{\rho}+\dt\rho,\varphi)\approx\frac{\tr(\varphi) \tilde{g}^3(\varphi)}{\cos{\tilde{\Theta}(\varphi)}}\sum_{k=0}^n\Delta J\br{t-\Delta t(\tr(\varphi))-k\tau(\tr(\varphi)),\tr(\varphi)+\dt r_k,\phi_o-\Delta\phi(\tr(\varphi))-k\delta(\tr(\varphi))},
\end{align}
where $\phi_o$ is the azimuth of the observer, while $\Delta\phi(\tr(\varphi))$ and $\Delta t(\tr(\varphi))$ respectively denote the $\O{k^0}$ pieces of the azimuthal angle and time accumulated along the photon's trajectory from disk to observer, which may be computed analytically using the asymptotic formulas \eqref{eq:Lapses}.

Plugging Eq.~\eqref{eq:InclinedIntensityFluctuations} into Eq.~\eqref{eq:InclinedC} yields
\begin{align}
	\label{eq:InclinedToyModel}
	\mathcal{C}(T,\varphi,\varphi')\approx\mathcal{N}(\varphi,\varphi')&\int\rho\ed(\dt\rho)\int\rho'\ed(\dt\rho')\sum_{k=0}^{n(\dt\rho)}\sum_{k'=0}^{n'(\dt\rho')}\,G_{\ell_r}\br{\tr(\varphi)-\tr(\varphi')+\dt r_{k}-\dt r_{k'}}\notag\\
	&\quad\times G_{\ell_t}\br{T+\Delta t(\tr(\varphi))-\Delta t(\tr(\varphi'))+k\tau(\tr(\varphi))-k'\tau(\tr(\varphi'))}\notag\\
	&\quad\times G_{\ell_\phi}^\circ\br{\Delta\phi(r(\tilde{\varphi}))-\Delta\phi(r(\tilde{\varphi}'))+k\delta\pa{\tr(\varphi)}-k'\delta\pa{\tr(\varphi')}},
\end{align}
where we introduced a prefactor
\begin{align}
	 \mathcal{N}(\varphi,\varphi')=\frac{\mathcal{J}\tr(\varphi)\tr(\varphi')\tilde{g}^3(\varphi)\tilde{g}^3(\varphi')}{\cos{\tilde{\Theta}(\varphi)}\cos{\tilde{\Theta}(\varphi')}}.
\end{align}
We wish to emphasize that the precise form of this prefactor depends on the specific assumptions of our toy model and more specifically, the motion of the emitters in the disk.  Nonetheless, the autocorrelation will still exhibit universal features.

As in Sec.~\ref{sec:PolarObserver} above, we have kept in Eq.~\eqref{eq:InclinedToyModel} only $\O{\dt\rho^0}$ terms in $\Delta\phi$ and $\Delta t$ and ignored subleading corrections.  In contrast to the polar case of Sec.~\ref{sec:PolarObserver}, however, here these terms are generically nontrivial functions of $\varphi$.  Following the same approach as before, we can interchange integration and summation in Eq.~\eqref{eq:InclinedToyModel}, and approximate near the photon shell $\rho\approx\tilde{\rho}(\varphi)$ and for large enough $k<n$,
\begin{align}
	\ed\pa{\dt\rho}\approx\iota(\varphi)e^{-k\gamma(\tr(\varphi))}\ed\pa{\dt r}_k,
\end{align}
where $\iota(\varphi)$ is computed in App.~\ref{app:Lensing} and given in Eq.~\eqref{eq:nu}.  Using Eq.~\eqref{eq:RadialIntegral} to integrate over radii, we obtain
\begin{align}
	\label{eq:InclinedToyModelIntermediateStep}
	\mathcal{C}(T,\varphi,\varphi')&\approx\mathcal{N}(\varphi,\varphi')\tilde{\rho}(\varphi)\tilde{\rho}(\varphi')\iota(\varphi)\iota(\varphi')W\Lambda\pa{\frac{\tr(\varphi)-\tr(\varphi')}{W}}\\
	&\quad\times\sum_{k=0}^\infty\sum_{k'=0}^\infty e^{-k\gamma(\tr(\varphi))}e^{-k'\gamma(\tr(\varphi'))}G_{\ell_t}\br{T+\Delta t(\tr(\varphi))-\Delta t(\tr(\varphi'))+k\tau(\tr(\varphi))-k'\tau(\tr(\varphi'))}\notag\\ 
	&\qquad\qquad\quad\times G_{\ell_\phi}^\circ\br{\Delta\phi(\tr(\varphi))-\Delta\phi(\tr(\varphi'))+k\delta\pa{\tr(\varphi)}-k'\delta\pa{\tr(\varphi')}},\notag
\end{align}
with $\Lambda(z)$ as given in Eq.~\eqref{eq:LambdaFunction}.  Equation~\eqref{eq:InclinedToyModelIntermediateStep} describes the intensity fluctuation 2PF on the photon ring for general inclination.  The argument of $\Lambda$ in Eq.~\eqref{eq:InclinedToyModelIntermediateStep} shows that, for nonzero BH spin, correlations between angles $\varphi$ and $\varphi'$ around the ring are significant (compared to correlations measured by a polar observer) only if the corresponding photon shell radii satisfy $\tr(\varphi)-\tr(\varphi')\lesssim W$.  One clear implication is that \textit{observations at small inclination are favorable} for measuring correlation around the ring at significant angular separation.  Note also that the relation \eqref{eq:ScreenCoordinates} defines $\tr(\varphi)$, but there is no closed-form expression for its inverse; thus, it could sometimes be more convenient to view the radius $\tr$ as a parameter along the ring in Eq.~\eqref{eq:InclinedToyModelIntermediateStep}, without direct reference to $\varphi$.  In the next section, however, we will approximate Eq.~\eqref{eq:InclinedToyModelIntermediateStep} for small observer inclinations by perturbatively inverting the relation $\tilde{r}(\varphi)$,  allowing us to directly express $\mathcal{C}(T,\varphi,\varphi')$ in terms of its arguments.

\section{Expansion in small inclination}
\label{sec:SmallInclination}

In the case of small observer inclination $0<\sin{\theta_o}\ll1$, we can significantly simplify Eq.~\eqref{eq:InclinedToyModelIntermediateStep}.  Expanding Eq.~\eqref{eq:ScreenCoordinates} to first order in inclination, we obtain a relation between the photon shell radii accessible to the observer and the angle on the sky, 
\begin{align}
	\label{eq:RadiusExpansion}
	\tr=\tr_0+\Xi a\sin{\theta_o}\cos{\varphi},
\end{align}
where, noting that $\tr_0^3=3M\tr_0^2-a^2\pa{\tr_0+M}$,
\begin{align}
	\Xi=\pa{\frac{\Delta(\tr_0)}{\tr_0-M}-M}\frac{4\tr_0^2\sqrt{\Delta(\tr_0)}}{3M^2\pa{\tr_0^2+a^2}+a^2\br{\Delta(\tr_0)-6M\tr_0}}.
\end{align}
Equation~\eqref{eq:RadiusExpansion} is useful for expanding many of the quantities appearing in Eq.~\eqref{eq:InclinedToyModelIntermediateStep}.  We can write
\begin{align}
	\label{eq:GenericExpansion}
	Y=Y_0+Y_1a\sin{\theta_o}\cos{\varphi},\qquad
	Y_1=\Xi\br{\pd_{\tr}Y}_{\tr=\tr_0},
\end{align}
for each of the quantities $Y\in\{\gamma,\delta,\tau,\tilde{g},\cos{\tilde{\Theta}}\}$.  The functions $\gamma(\tr)$, $\tau(\tr)$, $\delta(\tr)$ are explicitly given in Eq.~\eqref{eq:GeneralCriticalExponents}, while 
$g(\tr)$ is given in Eq.~\eqref{eq:Redshift}, and $\cos\Theta$ in Eq.~\eqref{eq:DirectionCosine}.  Here, we provide only an implicit definition of these expansion coefficients in order to avoid clutter, but it is straightforward to take the derivative and obtain their explicit, albeit quite lengthy, expressions.  Note that for low spins, $\tilde{g}_1\sim1/a$, so that the leading correction in small inclination to the redshift in Eq.~\eqref{eq:GenericExpansion} is $a$-independent.  Physically, the redshift is corrected at first order in inclination even for a nonrotating BH (with $a=0$), since radiation emitted from circular orbiters still exhibits Doppler shift even in the absence of BH spin.  This can be seen by expanding the general formula for $\tilde{\lambda}(\tr)$, which enters through the formula \eqref{eq:Redshift} for $g(\tilde{r})$, in small inclination,
\begin{align}
	\tilde{\lambda}\approx\frac{2\br{a^2(\tr_0+2M)-3M^2\tr_0}}{(\tr_0-M)^2}\Xi\sin{\theta_o}\cos{\varphi},
\end{align}
and noting that it has a finite, generically nonzero limit as $a\to0$.  We further note that the argument of $\Lambda$ in Eq.~\eqref{eq:InclinedToyModelIntermediateStep} is $\O{\sin{\theta_o}}$.  Since $\Lambda(z)=\Lambda(-z)$, to leading order in small $z$, $\Lambda(z)\approx\Lambda(0)+\O{z^2}$, and so $\Lambda$ does not admit corrections at $\O{\sin{\theta_o}}$. Using Eqs.~\eqref{eq:RadiusExpansion}, \eqref{eq:nu}, and \eqref{eq:Lapses}, we expand
\begin{align} \label{eq:small inclination expansion a}
	\Delta\phi(\tr(\varphi))-\Delta \phi(\tr(\varphi'))&=\Phi+\sin{\theta_o} \left[ a f^{e}_\phi\pa{\cos{\varphi}-\cos{\varphi'}}  + f^o_\phi\pa{\sin{\varphi}-\sin{\varphi'}} \right]+\O{\sin^2\theta_o},\\
	\Delta t(\tr(\varphi))-\Delta t(\tr(\varphi'))&=\sin{\theta_o} \left[ a f^{e}_t\pa{\cos{\varphi}-\cos{\varphi'}} + f^o_t\pa{\sin{\varphi}-\sin{\varphi'}} \right]+\O{\sin^2\theta_o},\label{eq:small inclination expansion b} \\
    \iota&=\iota_0+\sin{\theta_o} \left[ a \iota^{e}_1\cos{\varphi}  + \iota^o_1\sin{\varphi}  \right]+\O{\sin^2\theta_o},\label{eq:small inclination expansion c}
\end{align}
where
\begin{align} \label{eq:small inclination f, iota expansions}
	&\qquad f^{e}_\phi=\Xi\br{\pd_{\tr}\pa{\Delta\phi}}_{\tr=\tr_0},\qquad
	f^{e}_t=\Xi\br{\pd_{\tr}\pa{\Delta t}}_{\tr=\tr_0},\qquad
    \iota_1^{e}=\Xi\br{\pd_{\tr}\pa{\iota}}_{\tr=\tr_0}, \\ 
    &f^o_\phi=\left.\frac{\pd_{\sin\theta_o}\pa{\Delta\phi(\varphi)-\pa{\Delta\phi(-\varphi)}}}{{2\sin\varphi}}\right|_{\theta_o=0},\qquad
	f^{o}_t=\left.\frac{\pd_{\sin\theta_o}\pa{\Delta t(\varphi)-\pa{\Delta t(-\varphi)}}}{{2\sin\varphi}}\right|_{\theta_o=0},\notag \\
    & \qquad\qquad\qquad\qquad\qquad \iota_1^o=\left.\frac{\pd_{\sin\theta_o}\pa{\iota(\varphi)-\pa{\iota(-\varphi)}}}{{2\sin\varphi}}\right|_{\theta_o=0}, \notag 
\end{align}

At first order in inclination, the critical curve is still a perfect circle, though its center is horizontally shifted from the origin of the coordinate system $(\rho,\varphi)$, from $\rho=0$ to $\rho=\tilde{\rho}_0^2Xa\sin{\theta_o}$ with $\varphi=0$, where \cite{GrallaLupsasca2020c}
\begin{align}
	X=-\frac{2\pa{a^2\tilde{\rho}_0^2-27M^4-a^4}}{3\pa{M^2-a^2}\tilde{\rho}_0^4-2\pa{27M^4-30M^2a^2-a^4}\tilde{\rho}_0^2-96M^2a^4}.
\end{align}
We can therefore write, to first order in inclination,
\begin{align}
	\tilde{\rho}(\varphi)\approx\tilde{\rho}_0+\tilde{\rho}_0^2Xa\sin{\theta_o}\cos{\varphi}.
\end{align}
Putting everything together, and
defining $m=k-k'$ and $2s=k+k'-|m|$, Eq.~\eqref{eq:InclinedToyModelIntermediateStep} becomes, to $\O{\sin{\theta_o}}$,
\begin{align}
	\label{eq:SmallInclinationCorrelator}
	&\mathcal{C}(T,\Phi,\varphi+\varphi')\approx\\ 
	&c(\Phi,\varphi+\varphi')\sum_{m=-\infty}^\infty\sum_{s=0}^\infty e^{-(2s+|m|)\pa{\gamma_0+\gamma_1a\sin{\theta_o}\cos{\frac{\Phi}{2}}\cos{\frac{\varphi+\varphi'}{2}}}+m\gamma_1a\sin{\theta_o}\sin{\frac{\Phi}{2}}\sin{\frac{\varphi+\varphi'}{2}}}\notag\\
	&\times G_{\ell_t}\left[T+m\pa{\tau_0+\tau_1a\sin{\theta_o}\cos{\frac{\Phi}{2}}\cos{\frac{\varphi+\varphi'}{2}}} \right. \notag \\ &\left. ~~~~~~~~~~~~~~~~~~~~~~~~~~~~~~~~~~~~~+\sin{\theta_o}\sin{\frac{\Phi}{2}}\pa{a\pa{(2s+|m|)\tau_1+2f^e_t}\sin{\frac{\varphi+\varphi'}{2}}-2f^o_t\cos{\frac{\varphi+\varphi'}{2}}}\right]\notag\\
	&\times G_{\ell_\phi}^\circ\left[\Phi+m\pa{\delta_0+\delta_1a\sin{\theta_o}\cos{\frac{\Phi}{2}}\cos{\frac{\varphi+\varphi'}{2}}} \right. \notag \\ &\left. ~~~~~~~~~~~~~~~~~~~~~~~~~~~~~~~~~~~~~+\sin{\theta_o}\sin{\frac{\Phi}{2}}\pa{a\pa{(2s+|m|)\delta_1+2f^e_\phi}\sin{\frac{\varphi+\varphi'}{2}}-2f^o_\phi\cos{\frac{\varphi+\varphi'}{2}}}\right].\notag
\end{align}
where we introduced
\begin{align}
	&c(\Phi,\varphi+\varphi')=\pa{\frac{\iota_0\tilde{\rho}_0\tr_0\tilde{g}_0^3}{\cos{\tilde{\Theta}_0}}}^2\mathcal{J}W\Lambda(0)\\
	&\quad\times\left\{1+2\sin{\theta_o} \cos{\frac{\Phi}{2}} \br{\pa{3\frac{\tilde{g}_1}{\tilde{g}_0}+\frac{\iota^e_1}{\iota_0}+\frac{\Xi}{\tr_0}-\frac{\cos{\tilde{\Theta}_1}}{\cos{\tilde{\Theta}_0}}+\tilde{\rho}_0X} a \cos{\frac{\varphi+\varphi'}{2} } +\frac{\iota^o_1}{\iota_0}\sin\frac{\varphi+\varphi'}{2} } \right\}. \notag
\end{align}

If the conditions $(2s+|m|)\tau_1a\sin{\theta_o}\ll\ell_t$, $(2s+|m|)\delta_1a\sin{\theta_o}\ll\ell_\phi$, $f_ta\sin{\theta_o}\ll\ell_t$, and $f_\phi a\sin{\theta_o}\ll\ell_\phi$ are satisfied,\footnote{Formally, $s$ assumes arbitrarily large values, but since large-$s$ contributions are exponentially suppressed, we can safely use this approximation when we sum over all $s$ below.} then we can approximate Eq.~\eqref{eq:SmallInclinationCorrelator} using
\begin{align}
	G_{\ell_t}(z_0+\epsilon)\approx G_{\ell_t}(z_0)\pa{1-\frac{\epsilon z_0}{\ell_t^2}},\qquad
	G_{\ell_\phi}^\circ(z_0+\epsilon)\approx G_{\ell_\phi}^\circ(z_0)\pa{1-\frac{\epsilon\sin{z_0}}{\ell_\phi^2}}.
\end{align}
In turn, this allows us to explicitly perform the sum over $s$, resulting in
\begin{align}
	\mathcal{C}(T,\Phi,\varphi+\varphi') \approx\frac{c(\Phi,\varphi+\varphi')}{1-e^{-2\gamma_0}}&\sum_{m=-\infty}^\infty e^{-|m|\gamma_0+m\gamma_1a\sin{\theta_o}\sin{\frac{\Phi}{2}}\sin{\frac{\varphi+\varphi'}{2}}}\br{1-a\sin{\theta_o}b_m(T,\Phi,\varphi+\varphi')}\notag\\
	&\times G_{\ell_t}\br{T+m\tau_0+a\sin{\theta_o} m\tau_1\cos{\frac{\Phi}{2}}\cos{\frac{\varphi+\varphi'}{2}}}\notag\\
	&\times G_{\ell_\phi}^\circ\br{\Phi+m\delta_0+a\sin{\theta_o}m\delta_1\cos{\frac{\Phi}{2}}\cos{\frac{\varphi+\varphi'}{2}}},
\end{align}
where we introduced
\begin{align}
b_m(T,\Phi,\varphi+\varphi')&=
	\quad\cos\pa{\frac{\Phi}{2}}\cos\pa{\frac{\varphi+\varphi'}{2}}\gamma_1\pa{\frac{2}{e^{2\gamma_0}-1}+|m|}\\
&+\sin\pa{\frac{\Phi}{2}}\left\{\frac{T+m\tau_0}{\ell_t^2}\br{\Upsilon_t \sin\pa{\frac{\varphi+\varphi'}{2}}-\frac{2f^o_t}{a}\cos\pa{\frac{\varphi+\varphi'}{2}}}\right. \notag \\
&\left.+\frac{\sin\pa{\Phi+m\delta_0}}{\ell_\phi^2}\br{\Upsilon_\phi \sin\pa{\frac{\varphi+\varphi'}{2}} -\frac{2f^o_\phi}{a}\cos\pa{\frac{\varphi+\varphi'}{2}}}\right\},\notag\\
	\Upsilon_t&=\pa{\frac{2}{e^{2\gamma_0}-1}+|m|}\tau_1+2f^e_t,\qquad
	\Upsilon_\phi=\pa{\frac{2}{e^{2\gamma_0}-1}+|m|}\delta_1+2f^e_\phi.
\end{align}

An additional approximation can be made for low enough $m$, or more precisely, when the conditions $|m|\tau_1a\sin{\theta_o}\ll\ell_t$, $|m|\delta_1a\sin{\theta_o}\ll\ell_\phi$, and $|m|\gamma_1a{\sin{\theta_o}}\ll1$ are satisfied as well.  If we let $m_\mathrm{max}>0$ denote the maximal $m$ for which the above conditions hold, then the contribution of the $2m_\mathrm{max}+1$ peaks around $m=0$ to the 2PF is given by
\begin{align}
	\label{eq:SmallInclinationCorrelatorIntermediateStep}
	\mathcal{C}(T,\Phi,\varphi+\varphi') \approx\frac{c(\Phi,\varphi+\varphi')}{1-e^{-2\gamma_0}}\sum_{m=-m_\mathrm{max}}^{m_\mathrm{max}}& e^{-|m|\gamma_0}\br{1-a\sin{\theta_o}B_m(T,\Phi,\varphi+\varphi')}\notag\\
	&\times G_{\ell_t}\pa{T+m\tau_0}G_{\ell_\phi}^\circ\pa{\Phi+m\delta_0},
\end{align}
where
\begin{align}
	B_m(T,\Phi,\varphi+\varphi')&=b_m(T,\Phi,\varphi+\varphi')- m\gamma_1\sin\pa{\frac{\Phi}{2}}\sin\pa{\frac{\varphi+\varphi'}{2}}\\
	&\quad+m\cos\pa{\frac{\Phi}{2}}\cos\pa{\frac{\varphi+\varphi'}{2}}\br{\frac{T+m\tau_0}{\ell_t^2}\tau_1+\frac{\sin\pa{\Phi+m\delta_0}}{\ell_\phi^2}\delta_1}.\notag
\end{align}
It is natural to integrate over $\varphi+\varphi'$ to obtain an ``effective'' 2D correlator that can be compared to the one obtained for a polar observer in Sec.~\ref{sec:PolarObserver}.  More explicitly, we would like to compute
\begin{align}
	\label{eq:Cbar}
	\bar{\mathcal{C}}(T,\Phi)=\frac{1}{2\pi}\int\ed\pa{\frac{\varphi+\varphi'}{2}}\mathcal{C}(T,\Phi,\varphi+\varphi').
\end{align}
This quantity is trivial to first order in small inclination since all the pieces of Eq.~\eqref{eq:SmallInclinationCorrelatorIntermediateStep} that depend nontrivially on $\varphi+\varphi'$---which are also $\O{\sin{\theta_o}}$---integrate to zero.  Therefore, Eq.~\eqref{eq:Cbar} integrates precisely to Eq.~\eqref{eq:ToyModelIntermediateStep2} to first order in inclination, inclusive, and is corrected only at $\O{\sin^2{\theta_o}}$.  This is an encouraging fact from an observational perspective, since it means that for small inclinations, the results of Sec.~\ref{sec:PolarObserver} are robust against changes of inclination once $\varphi+\varphi'$ is integrated out.

\section{Observational prospects and considerations}
\label{sec:Observations}

We conclude by briefly commenting on the observational prospects for measuring $\mathcal{C}(T,\varphi,\varphi')$.  Observing several clearly separated maxima in $\mathcal{C}(T,\varphi,\varphi')$ would provide strong evidence that some of the arriving photons were very strongly lensed by the BH.  Such an observation could also provide a test of General Relativity, since the theory makes a universal prediction for the self-similar structure, locations, and relative heights of these peaks.  
Deviations from General Relativity can modify our results for $\mathcal{C}(T,\varphi,\varphi')$ in many different ways; to fully understand the effects of such potential deviations on the fluid evolution and ray-tracing would require significant work beyond the scope of this paper.  Assuming General Relativity, the 2PF on the ring could be used to estimate \textit{both the mass and spin of the BH}, as well as statistical properties of the accretion flow.  A successful measurement would not require specialized emission conditions, such as a compact hot spot; we only require that the image be variable.  Because sources such as Sgr~A* and M87* are known to exhibit horizon-scale variability \cite{Fish2011,Johnson2015,Gravity2018,EHT2019c,EHT2019d}, measuring $\mathcal{C}(T,\varphi,\varphi')$ is simply a question of achieving the required sensitivity.  We will now derive rough estimates for the sensitivity requirements.

We will first consider an observation in which images are measured perfectly, with unlimited angular and temporal resolution.  In this case, the required sensitivity is determined solely by the source stochasticity.  An observation that continuously spans a timescale of $t_\mathrm{obs}$ will sample $N_t\sim t_\mathrm{obs}/\ell_t$ stochastic realizations in time and $N_\phi\sim 2\pi/\ell_\phi$ stochastic realizations in angle.  The $m^\text{th}$ correlation peak will have an amplitude that depends on the fraction $f_{\rm img}$ of the image flux that is fluctuating, the partial reduction in correlation amplitude from radial averaging, and reductions for the finite-delay envelope along null geodesics passing through the emitting region.  We can approximate the magnitude of the (dimensionless) correlation peak as $\mathcal{C}(T,\Phi)/I_0^2 \sim e^{-|m|\gamma_0}f_\mathrm{img}^2$, where $I_0$ is the average flux density of the ring. 

Given $N$ independently sampled pairs of image intensity with respective lags $(T,\Phi)$, the signal-to-noise ratio (SNR) for their correlation is approximately $N^{1/2}|\mathcal{C}(T,\Phi)/\mathcal{C}(0,0)|$.  Hence, even in the idealized case of infinite resolution, there will be a finite SNR at a fixed angle,
\begin{align}
	\mathrm{SNR}_\infty(\phi)\sim e^{-|m|\gamma_0}\sqrt{\frac{t_\mathrm{obs}}{\ell_t}},
\end{align}
while the SNR when combining information from all angles is 
\begin{align}
	\mathrm{SNR}_\infty\sim e^{-|m|\gamma_0}\sqrt{\frac{2\pi t_\mathrm{obs}}{\ell_\phi\ell_t}}.
\end{align}

A real observation will also have limitations from finite angular resolution $\theta_\mathrm{obs}$, limited temporal sampling cadence $\Delta t_\mathrm{obs}$, and additional image noise that is related to details of the instrument and observation (e.g., finite baseline coverage or calibration uncertainties for an interferometer).  Finite angular resolution is likely to be the most significant of these limitations; if the resolution is insufficient to resolve $\ell_\phi$, then it will reduce the measured correlation and will decrease the number of statistical realizations that can be combined to improve sensitivity.  Letting $\theta_\mathrm{ph}$ denote the angular radius of the photon ring, the image azimuthal resolution is $\theta_\mathrm{obs}/\theta_\mathrm{ph}\geq\ell_\phi$, and the SNR becomes
\begin{align}
	\mathrm{SNR}&\sim e^{-|m| \gamma_0}\sqrt{\frac{2\pi t_\mathrm{obs}\theta_\mathrm{ph}^2\ell_\phi}{\theta_\mathrm{obs}^2\ell_t}}\\
	&=\frac{\ell_\phi\theta_\mathrm{ph}}{\theta_\mathrm{obs}}\mathrm{SNR}_\infty.
\end{align}
The effects of coarse temporal sampling $\Delta t_\mathrm{obs}\geq\ell_t$ are identical: $\mathrm{SNR}\propto\frac{\ell_t}{\Delta t_\mathrm{obs}}\mathrm{SNR}_\infty$.  Finally, errors in a reconstructed image will have nontrivial correlation structure across the image (e.g., related to systematic calibration errors) and across time (e.g., related to limited baseline coverage that may be the same in different observing epochs).  We can crudely represent these as a source of added stochastic noise, which effectively serves to reduce the correlation by some factor $f_\mathrm{obs} \sim f_\mathrm{img}^2$.  Hence, we expect it to be more difficult to detect the correlation structure in the case of weak image fluctuations, even if the normalized correlation is large.

M87* is a natural target to consider.  The EHT has already demonstrated that daily horizon-scale snapshot images of M87* are feasible, and has already measured horizon-scale intrinsic variability on a timescale of ${\sim}10GM/c^3$ \cite{EHT2019c,EHT2019d}. Significant variability is also seen in longer-wavelength monitoring \cite{Walker2018}. 
Moreover, M87* anchors a prominent kpc-scale jet; measurements of the jet/counterjet brightness ratio, kinematics, and differential limb brightening are all consistent with a viewing inclination of $17^\circ \pm 3^\circ$ \cite{Walker2018}. Thus, under the assumption that the black hole spin vector is aligned with the jet, the black hole in M87* is viewed at small inclination and the approximations in sec.~\ref{sec:SmallInclination} are appropriate.   
This supermassive BH has $M/D\sim4\,\mu{\rm as}$ and $GM/c^3\approx9\,{\rm hours}$ \cite{EHT2019f}, while the EHT currently has an angular resolution of ${\sim}20\,\mu{\rm as}$.  Thus, $\theta_\mathrm{obs}/\theta_\mathrm{ph}\sim10\ell_\phi$.  We then obtain, for the $m=1$ peak in the autocorrelation,
\begin{gather}
	\text{M87*:}\notag\\
	\mathrm{SNR}_\infty\sim0.7
	\pa{\frac{e^{-\gamma_0}}{e^{-\pi}}}
	\pa{\frac{\ell_\phi}{5^\circ}}^{-\frac{1}{2}} 
	\pa{\frac{\ell_t}{1\,{\rm day}}}^{-\frac{1}{2}}
	\pa{\frac{f_\mathrm{obs}}{0.1}} 
	\pa{\frac{t_\mathrm{obs}}{1\,{\rm year}}}^{\frac{1}{2}},\\ 
	\mathrm{SNR}\sim0.02
	\pa{\frac{e^{-\gamma_0}}{e^{-\pi}}}
	\pa{\frac{\ell_\phi}{5^\circ}}^\frac{1}{2}
	\pa{\frac{\ell_t}{1\,{\rm day}}}^\frac{1}{2}
	\pa{\frac{f_\mathrm{obs}}{0.1}}
	\pa{\frac{t_\mathrm{obs}}{1\,{\rm year}}}^{\frac{1}{2}} 
	\pa{\frac{\theta_\mathrm{obs}}{20\,\mu{\rm as}}}^{-1} 
	\pa{\frac{\Delta t_\mathrm{obs}}{3\,{\rm days}}}^{-1}.\notag
\end{gather}
Here, we set an optimistic characteristic value of 3\,days for $\Delta t_\mathrm{obs}$, which accounts for limitations that may impede continuous observations, such as poor weather.  Thus, even with a perfect instrument and continuous monitoring, detecting a strong signal in $\mathcal{C}$ for M87* will likely require some combination of rapid BH spin (increasing $e^{-\gamma_0}$ by a factor of ${\sim}2$), a high fraction of the image that is variable, and many months or years of observation.  Current observations lose another factor of ${\sim}35$ in sensitivity from the combination of limited sampling in time (${\sim}3{\times}$) and the limited image resolution (${\sim}10{\times}$).  Nevertheless, observations of M87* with the EHT every few days over a span of a few months or years would allow first estimates of $\mathcal{C}$.

A second possible target for observations is the Galactic Center supermassive BH, Sgr~A*.  This supermassive BH has $M/D\sim5\,\mu{\rm as}$ and $GM/c^3\approx21\,{\rm seconds}$ \cite{Do2019,Gravity2020}. While the mass and distance are tightly constrained, Sgr~A* does not have an observed jet, and the the black hole inclination has no firm constraints. If the inclination of Sgr~A* is large, our estimate for the SNR may need to be significantly modified. As with M87*, Sgr~A* shows ${\sim}20\%$ variability in its total flux density \cite{Bower2015}, thereby implying significant image variability.  
In addition, the gravitational timescale is shorter than Earth rotation timescales. Nevertheless, extensions of the EHT that are sufficient to reconstruct movies of Sgr~A* would allow estimates of $\mathcal{C}$ with continuous observations \cite{Johnson2017,Bouman2018,Palumbo2019,Blackburn2019}.  We then obtain
\begin{gather}
	\text{Sgr~A*:}\notag\\
	\mathrm{SNR}_\infty\sim30
	\pa{\frac{e^{-\gamma_0}}{e^{-\pi}}}
	\pa{\frac{\ell_\phi}{5^\circ}}^{-\frac{1}{2}}
	\pa{\frac{\ell_t}{1\,{\rm minute}}}^{-\frac{1}{2}}
	\pa{\frac{f_\mathrm{obs}}{0.1}}
	\pa{\frac{t_\mathrm{obs}}{1\,{\rm year}}}^\frac{1}{2},\\
	\mathrm{SNR}\sim0.6
	\pa{\frac{e^{-\gamma_0}}{e^{-\pi}}}
	\pa{\frac{\ell_\phi}{5^\circ}}^\frac{1}{2} 
	\pa{\frac{\ell_t}{1\,{\rm minute}}}^\frac{1}{2}
	\pa{\frac{f_\mathrm{obs}}{0.1}}
	\pa{\frac{t_\mathrm{obs}}{1\,{\rm year}}}^\frac{1}{2}
	\pa{\frac{\theta_\mathrm{obs}}{20\,\mu{\rm as}}}^{-1} 
	\pa{\frac{\Delta t_\mathrm{obs}}{5\,{\rm minutes}}}^{-1} 
	.\notag
\end{gather}
Here, we set an optimistic characteristic value of $5\,{\rm minutes}$ for $\Delta t_\mathrm{obs}$.  While the $\Delta t_\mathrm{obs}$ for M87* depends on logistics related to conducting observations, $\Delta t_\mathrm{obs}$ for Sgr~A* instead corresponds to the minimum time required to form an image.  The expected SNR for Sgr~A* is significantly higher than M87* for the same observing duration, primarily because of the significantly shorter coherence timescales expected: $M_\text{M87*}/M_\text{SgrA*}\approx1500$.

In both cases, we can quantify the improvement afforded by using imaging rather than analysis of the ``light-curve'' measured for an unresolved source.  The latter has the advantage of requiring only one telescope rather than an entire interferometric array.  However, the SNR is significantly higher with imaging than with a light-curve analysis, $\mathrm{SNR}^\mathrm{LC}$:
\begin{align}
	\frac{\mathrm{SNR}_\infty}{\mathrm{SNR}_\infty^\mathrm{LC}}&\sim\frac{2\pi}{\ell_\phi}
	\approx70,\\
	\frac{\mathrm{SNR}}{\mathrm{SNR}^\mathrm{LC}}&\sim\frac{2\pi \theta_\mathrm{ph}}{\theta_\mathrm{obs}}
	\approx 6.
\end{align}
Thus, image analysis with the resolution of the EHT decreases the required observing time by a factor of ${\sim}40$, while observing with a significantly enhanced array would decrease the required observing time by a factor of ${\sim}\,5\times10^3$.  An image analysis also has the advantage of identifying azimuthal structure in the correlation function, which provides information about the BH spin.  However, detecting signatures of the $m=2$ and higher-order correlation peaks is unlikely with our approach; even with an image-based analysis and perfect resolution for Sgr~A*, a significant detection of $m=2$ signatures would require many years of continuous observations.  Studies of unusual events, such as flares from compact emission regions, are more likely to yield the requisite SNR. 

A separate issue for detecting $\mathcal{C}$ is to distinguish the universal correlation structure that reflects properties of the photon shell from the astrophysical correlation structure that reflects properties of the emitting plasma (see, e.g., Fig.~\ref{fig:TimeAutocorrelation}).  In particular, a plasma with relativistic rotation may have significant nontrivial correlation structures even at the same values $(T,\Phi)$ at which the lensing gives peaks.  Nevertheless, we expect this contamination to be insignificant when the astrophysical correlation is small at the location of the lensing peaks because the correlations from astrophysics and lensing are approximately additive.  
However, if the astrophysical correlation is comparable to the correlation from lensing, then the lensing signature will be much more difficult to detect.

Although we have only considered fluctuations in the Stokes intensity $I$, a similar universal structure will be imprinted in the 2PF of observed fluctuations in the other Stokes parameters $Q$, $U$ and $V$.  However, while the image correlation structure will be governed by the same critical exponents as $I$ (determined solely by the achromatic lensing of the BH), the correlation structure of the emissivity may differ because changes in the magnetic field direction affect polarization differently than total intensity.  Moreover, the fluctuations may highlight differential Faraday effects among the different interfering paths, which can potentially be quite strong for sources such as M87* and Sgr~A*, even when the source is optically thin \cite{Moscibrodzka2017,Rosales2018,Ricarte2020}.  It would be interesting (and likely difficult) to understand the correlation structure of source fluctuations in the linear polarization $Q$ and $U$, and we defer this question to future work.

\section*{Acknowledgements}

We thank Charles Gammie for helpful suggestions and corrections.  SH and AL gratefully acknowledge support from the Jacob Goldfield Foundation.  GNW was supported by a Donald C.~and F.~Shirley Jones Fellowship.  M.~D.~J. acknowledges support from the National Science Foundation (AST-1716536, AST-1935980) and the Gordon and Betty Moore Foundation (GBMF-5278).

\appendix

\section{Lensed images of an equatorial disk}
\label{app:Lensing}

In this appendix, we use the analytic techniques developed in Ref.~\cite{GrallaLupsasca2020a,GrallaLupsasca2020b} to present a proof of Eq.~\eqref{eq:RingDemagnification} and derive an expression for the quantity $\nu$ introduced therein.  The simplest approach relies on the method of matched asymptotic expansions, which was applied to near-critical geodesics in App.~B of Ref.~\cite{GrallaLupsasca2020a}.  An alternative method would be to use the exact solution of the null geodesic equation in Kerr obtained in Ref.~\cite{GrallaLupsasca2020b} and expand it near criticality.  We will first employ the simple approach and then briefly explain how one can confirm that it agrees with the second.

\subsection{Matched asymptotic expansion for near-critical geodesics}

The $(r,\theta)$ component of the null geodesic equation in Kerr can be recast in integral form as
\begin{gather}
	\label{eq:NullGeodesics}
	I_r=\fint_{r_s}^{r_o}\frac{\ed r}{\pm_r\sqrt{\mathcal{R}(r)}}
	=\fint_{\theta_s}^{\theta_o}\frac{\ed\theta}{\pm_\theta\sqrt{\Theta(\theta)}}
	=G_\theta,\\
	\mathcal{R}(r)=\br{\pa{r^2+a^2}-a\lambda}^2-\Delta(r)\br{\eta+\pa{\lambda-a}^2},\qquad
	\Theta(\theta)=\eta+a^2\cos^2{\theta}-\lambda^2\cot^2{\theta},
\end{gather}
with the slashed integrals indicating that the signs $\pm_{r,\theta}$ switch at every radial and polar turning point, respectively.  In this paper, we consider photons received by a fixed observer at inclination $\theta_o$ and large radius $r_o\to\infty$ after being emitted from radius on the surface of an equatorial disk $\theta_s=\pi/2$.  We wish to invert Eq.~\eqref{eq:NullGeodesics} in order to obtain the radial trajectory $r_s(G_\theta)$ as a function of the conserved quantities $(\lambda,\eta)$ parametrizing the observer sky via Eq.~\eqref{eq:ScreenCoordinates}.  This can be done exactly, but in this section we will obtain a simplified formula for $r_s$ that holds near the photon shell for near-critical geodesics with conserved quantities $(\lambda,\eta)$ close to the critical values \eqref{eq:CriticalGeodesics}.  Such geodesics must appear very close to the critical curve and may be parametrized by $(\tr,d)$, where $0<\ab{d}\ll1$ denotes their small perpendicular distance from the closest point $(\tilde{\rho}(\tr),\tilde{\varphi}(\tr))$ on the critical curve (see Fig.~3 of Ref.~\cite{GrallaLupsasca2020a}).

As an intermediate step in our calculation, we follow App.~B of Ref.~\cite{GrallaLupsasca2020a} and replace $d$ by a quantity $\dt r_0$ defined in terms of Bardeen's coordinates $(\alpha,\beta)=(\rho\cos{\varphi},\rho\sin{\varphi})$ as
\begin{align}
	\label{eq:PerpendicularDistance}
	d=\frac{2\tr^4\tilde{\chi}}{\Delta(\tr)}\frac{\dt r_0^2}{\sqrt{\tilde{\beta}^2+\tilde{\psi}^2}},\qquad
	\tilde{\psi}=\tilde{\alpha}-\pa{\frac{\tr+M}{\tr-M}}a\sin{\theta_o},\qquad
	\tilde{\chi}=1-\frac{M\Delta(\tr)}{\tr\pa{\tr-M}^2}.
\end{align}
For light rays outside the critical curve with $d\propto\dt r_0^2>0$, the radius of closest approach to the BH is $r_\mathrm{min}=\tr\pa{1+\dt r_0}$.  A near-critical light ray traced backwards from a position outside/inside the critical curve in the observer sky to a radius $r_s=\tr\pa{1+\dt r_s}$ near the photon shell has a radial geodesic integral $I_r$ given by Eqs.~(B30) and (B32) in Ref.~\cite{GrallaLupsasca2020a}, respectively,
\begin{align}
	I_r^\mathrm{nf,out}(\dt r_s,\infty)&=-\frac{1}{2\tr\sqrt{\tilde{\chi}}}\br{\arctanh\sqrt{\tilde{\chi}}+\arctanh\pa{\frac{\sqrt{\dt r_s^2-\dt r_0^2}}{\dt r_s}}+\frac{1}{2}\log\pa{\frac{1-\tilde{\chi}}{\pa{8\tilde{\chi}}^2}\dt r_0^2}},\\
	I_r^\mathrm{nf,in}(\dt r_s,\infty)&=-\frac{1}{2\tr\sqrt{\tilde{\chi}}}\br{\arctanh\sqrt{\tilde{\chi}}+\arctanh\pa{\frac{\dt r_s}{\sqrt{\dt r_s^2-\dt r_0^2}}}+\frac{1}{2}\log\pa{\frac{1-\tilde{\chi}}{\pa{8\tilde{\chi}}^2}\ab{\dt r_0^2}}},
\end{align}
where we took $\dt r_b\to\infty$ in both expressions since we only consider far observers.  Solving $I_r=G_\theta$ for the portion of the trajectory in the photon shell results in
\begin{align}
	r_s=\tr\pa{1+\dt r_s^\mathrm{in/out}},\qquad
	\dt r_s^\mathrm{out}=\dt r_0\cosh{\tau},\qquad
	\dt r_s^\mathrm{in}=-\sqrt{-\dt r_0^2}\sinh{\tau},
\end{align}
with
\begin{align}
	\tau=2\tr\sqrt{\tilde{\chi}}G_\theta+\arctanh\sqrt{\tilde{\chi}}+\frac{1}{2}\log\pa{\frac{1-\tilde{\chi}}{\pa{8\tilde{\chi}}^2}\dt r_0^2}.
\end{align}
Expanding the hyperbolic functions in small $0<\dt r_0^2\ll1$ results in the single expression
\begin{align}
	\dt r_s=\dt r_s^\mathrm{in/out}
	=\frac{1+\sqrt{\tilde{\chi}}}{16\tilde{\chi}}\dt r_0^2e^{2\tr\sqrt{\tilde{\chi}}G_\theta}+\frac{4\tilde{\chi}}{1+\sqrt{\tilde{\chi}}}e^{-2\tr\sqrt{\tilde{\chi}}G_\theta}.
\end{align}
Finally, plugging in for $d$ yields
\begin{align}
	r_s(G_\theta)=\tr\br{1+\frac{1+\sqrt{\tilde{\chi}}}{32\tr^4\tilde{\chi}^2}\sqrt{\tilde{\beta}^2+\tilde{\psi}^2}\Delta(\tr)de^{2\tr\sqrt{\tilde{\chi}}G_\theta}+\frac{4\tilde{\chi}}{1+\sqrt{\tilde{\chi}}}e^{-2\tr\sqrt{\tilde{\chi}}G_\theta}}.
\end{align}
This formula describes the portion of a near-critical geodesic's radial motion near the photon shell, as a function of its angular geodesic integral $G_\theta$ (equal to the elapsed Mino time) and position $(\tr,d)$ near the critical curve in the observer sky.

In the case of a polar observer, Eq.~(78) of Ref.~\cite{GrallaLupsasca2020a} implies that such a geodesic intersects the equatorial plane whenever [recalling Eq.~\eqref{eq:CriticalCurveRadius} for the definition of $\tilde{\rho}_0$]
\begin{align}
	\label{eq:PolarObserverAngularIntegral}
	G_\theta=\frac{2m+1}{4\tr_0\sqrt{\tilde{\chi}}}\gamma_0,\qquad
	\gamma_0=\frac{4\tr_0\sqrt{\tilde\chi}}{\sqrt{\tilde{\rho}_0^2-a^2}}K\pa{\frac{a^2}{a^2-\tilde{\rho}_0^2}},
\end{align}
for some integer $m$ labeling equatorial crossings.  As such, noting that for a polar observer, $\tilde{\psi}=\tilde{\alpha}$ and therefore $\tilde{\beta}^2+\tilde{\psi}^2=\tilde{\rho}_0^2$, the radius of the $m^\text{th}$ equatorial crossing is
\begin{align}
	r_s^{(m)}=\tr\br{1+\frac{1+\sqrt{\tilde{\chi}}}{32\tr^4\tilde{\chi}^2}\tilde{\rho}_0\Delta(\tr)de^{\pa{m+\frac{1}{2}}\gamma_0}+\frac{4\tilde{\chi}}{1+\sqrt{\tilde{\chi}}}e^{-\pa{m+\frac{1}{2}}\gamma_0}}.
\end{align}

More generally, for an equatorial source but observer at general inclination $\theta_o>0$, we have from Eqs.~(20) and (43) of Ref.~\cite{GrallaLupsasca2020a}
\begin{align}
	\label{eq:InclinedObserverAngularIntegral}
	G_\theta=\frac{2m}{4\tr\sqrt{\tilde{\chi}}}\gamma\mp_of_o,\qquad
	\gamma=\frac{4\tr\sqrt{\tilde{\chi}}}{a\sqrt{-\tilde{u}_-}}\tilde{K},
\end{align}
where $\pm_o=\sign{\beta}$ and using Eq.~(12) of Ref.~\cite{GrallaLupsasca2020a}, 
\begin{align}
	f_o=\frac{1}{a\sqrt{-\tilde{u}_-}}F\pa{\arcsin\pa{\frac{\cos{\theta_o}}{\sqrt{\tilde{u}_+}}}\left|\frac{\tilde{u}_+}{\tilde{u}_-}\right.}.
\end{align}
Hence, we obtain the general expression
\begin{align}
	\label{eq:EquatorialCrossing}
	r_s^{(m)}=\tr\br{1+\frac{1+\sqrt{\tilde{\chi}}}{32\tr^4\tilde{\chi}^2}\sqrt{\tilde{\beta}^2+\tilde{\psi}^2}\Delta(\tr)de^{\mp_o2\tr\sqrt{\tilde{\chi}}f_o}e^{m\gamma}+\frac{4\tilde{\chi}}{1+\sqrt{\tilde{\chi}}}e^{\pm_o2\tr\sqrt{\tilde{\chi}}f_o}e^{-m\gamma}}.
\end{align}
We define the polar tangential angle $\tilde{\Psi}(\varphi)$ to the critical curve as the angle $0\le\tilde{\Psi}\le\pi$ such that
\begin{align}
	\tan{\tilde{\Psi}(\varphi)}=\frac{\tilde{\rho}(\varphi)}{\tilde{\rho}'(\varphi)}.
\end{align}
This angle is such that $d=\dt\rho\sin{\tilde{\Psi}}$, and so we obtain the following general relation for polar curves:
\begin{align}
	\frac{\pd}{\pd(\dt\rho)}=\sin\br{\tilde{\Psi}(\varphi)}\frac{\pd}{\pd d}.
\end{align}
This implies that
\begin{align}
	\iota\frac{\pd r_s^{(m)}}{\pd(\dt\rho)}=e^{m\gamma}+\O{e^{-m\gamma}},
\end{align}
where
\begin{align}
\label{eq:nu}
	\frac{1}{\iota}=\frac{1+\sqrt{\tilde{\chi}}}{32\tr^3\tilde{\chi}^2}\sqrt{\tilde{\beta}^2+\tilde{\psi}^2}\Delta(\tr)e^{\mp_o2\tr\sqrt{\tilde{\chi}}f_o}\sin\br{\tilde{\Psi}(\varphi)},
\end{align}
a fact we use in our computation of the 2-point correlator of intensity fluctuations in Sec.~\ref{sec:PolarObserver} above.

\subsection{Near-critical expansion of the exact transfer function}

Recently \cite{GrallaLupsasca2020b}, the null geodesic equation in the Kerr spacetime was completely solved analytically in terms of elliptic functions.  Using these results, Ref.~\cite{GrallaLupsasca2020a} showed that the radial trajectory of a light ray shot backwards into the geometry from radial infinity is exactly given by
\begin{align}
	\label{eq:TransferFunction}
	r_s(\tau)=\frac{r_4r_{31}-r_3r_{41}\sn^2\pa{\frac{1}{2}\sqrt{r_{31}r_{42}}\tau-\mathcal{F}_o\big|k}}{r_{31}-r_{41}\sn^2\pa{\frac{1}{2}\sqrt{r_{31}r_{42}}\tau-\mathcal{F}_o\big|k}},
\end{align}
with $\tau$ denoting the Mino time along the trajectory from emission point to observer, and
\begin{align}
	\mathcal{F}_o=F\pa{\left.\arcsin{\sqrt{\frac{r_{31}}{r_{41}}}}\right|k},\quad
	k=\frac{r_{32}r_{41}}{r_{31}r_{42}}.
\end{align}
Here, we introduced the notation
\begin{align}
	r_{ij}=r_i-r_j,
\end{align}
with $\cu{r_1,r_2,r_3,r_4}$ denoting the roots of the quartic potential $\mathcal{R}(r)$ appearing in the radial geodesic integral $I_r$.  Analytic expressions for these roots are derived in Ref.~\cite{GrallaLupsasca2020b}.

Plugging in $\tau=G_\theta$, with $G_\theta$ as given in Eq.~\eqref{eq:PolarObserverAngularIntegral} or Eq.~\eqref{eq:InclinedObserverAngularIntegral} (according to whether the observer is polar or inclined, respectively) defines the exact transfer functions describing the optical appearance of an equatorial disk,
\begin{align}
	r_s^{(m)}(\rho,\varphi)=r_s\pa{G_\theta^m(\rho,\varphi)}.
\end{align}
Contours of the transfer functions for the direct and first lensed image of the disk are shown in Fig.~6 of Ref.~\cite{GrallaLupsasca2020a} and in Figs. 3 and 4 of Ref.~\cite{Gates2020a}.

It is possible to derive the formula \eqref{eq:EquatorialCrossing} for the radius of equatorial crossings by expanding the exact transfer function \eqref{eq:TransferFunction} near the critical curve.  For near-critical light rays with small $0<\dt r_0^2\ll1$, the radial roots $\cu{r_1,r_2,r_3,r_4}$ are approximately
\begin{align}
	r_1&=\tr\pa{-1-2\sqrt{1-\tilde{\chi}}}+\O{\dt r_0^2},\\
	r_2&=\tr\pa{-1+2\sqrt{1-\tilde{\chi}}}+\O{\dt r_0^2},\\
	r_3&=\tr\pa{1-\dt r_0}+\O{\dt r_0^2},\\
	r_4&=\tr\pa{1+\dt r_0}+\O{\dt r_0^2}.
\end{align}
Plugging these expressions into Eq.~\eqref{eq:TransferFunction} and expanding in small $\dt r_0$ reproduces the formula \eqref{eq:EquatorialCrossing}.  However, this expansion is difficult and crucially requires the use of a nontrivial asymptotic expansion of the incomplete elliptic integral of the first kind, derived in App.~C of Ref.~\cite{Gates2020a},
\begin{align}
	F\pa{\left.\frac{\pi}{2}-\epsilon\right|1-A\epsilon^2}\stackrel{\epsilon\to0}{\approx}-\frac{1}{2}\log\pa{A\epsilon^2}+\log{4}-\log\pa{\frac{1}{\sqrt{A}}+\sqrt{1+\frac{1}{A}}}.
\end{align}

\subsection{Time lapse and azimuth swept}

In general, the time elapsed and azimuth swept along a light ray are given by
\begin{align}
	\Delta t=t_o-t_s&=I_t+a^2G_t,\\
	\Delta\phi=\phi_o-\phi_s&=I_\phi+\lambda G_\phi,
\end{align}
with the geodesic integrals $I_t$, $I_\phi$, $G_t$ and $G_\phi$ defined in Eqs.~(13) of Ref.~\cite{GrallaLupsasca2020b}.  The angular geodesic integrals $G_t$ and $G_\phi$ do not simplify near criticality, and their general expression for near-critical rays shot back into the geometry is given in Sec.~IIA of Ref.~\cite{GrallaLupsasca2020a}.  On the other hand, the radial geodesic integrals $I_t$ and $I_\phi$ may be computed by the method of matched asymptotic expansions, resulting in the simplified formulas given in Eqs.~(B34) and (B36) of Ref.~\cite{GrallaLupsasca2020a},
\begin{align}
	I_t&=\tr^2\pa{\frac{\tr+3M}{\tr-M}}I_r-\frac{\tr}{2\sqrt{\tilde{\chi}}}\br{\mathcal{Q}_t(\dt r_o)-\mathcal{Q}_t(0)},\\
	I_\phi&=a\pa{\frac{\tr+M}{\tr-M}}I_r-\frac{aM}{\tr\sqrt{\tilde{\chi}}}\br{\mathcal{Q}_\phi(\dt r_o)-\mathcal{Q}_\phi(0)},
\end{align}
where $\dt r_o\to\infty$ for a distant observer and the auxiliary quantities $\mathcal{Q}_t$ and $\mathcal{Q}_\phi$ are defined in App.~B of Ref.~\cite{GrallaLupsasca2020a}.  Using $I_r=G_\theta$, this implies that
\begin{subequations}
\label{eq:Lapses}
\begin{align}
	\Delta t&=\tr^2\pa{\frac{\tr+3M}{\tr-M}}G_\theta+a^2G_t-\frac{\tr}{2\sqrt{\tilde{\chi}}}\br{\mathcal{Q}_t(\dt r_o)-\mathcal{Q}_t(0)},\\
	\Delta\phi&=a\pa{\frac{\tr+M}{\tr-M}}G_\theta+\lambda G_\phi-\frac{aM}{\tr\sqrt{\tilde{\chi}}}\br{\mathcal{Q}_\phi(\dt r_o)-\mathcal{Q}_\phi(0)}. 
\end{align}
\end{subequations}
The $d$-dependence drops out entirely at leading order.  While the azimuth swept $\Delta\phi$ remains finite in the limit $\dt r_o\to\infty$, the time elapsed along the trajectory from photon shell to observer naturally diverges linearly as $\dt r_o\to\infty$.  However, for the computations in this paper, we only care about the difference in time delay between received signals, so we may replace $\Delta t$ for a light ray by its difference with the time elapsed along a reference trajectory, such as the one providing the first image of any given bulk fluctuation.

\section{Redshift factor for emission from an equatorial disk}
\label{app:Redshift}

The Kerr geometry only admits stable circular orbits down to the radius of the Innermost Stable Circular Orbit (ISCO) at
\begin{subequations}
\label{eq:ISCO}
\begin{gather}
	r_\mathrm{ms}=M\pa{3+Z_2-\sqrt{\pa{3-Z_1}\pa{3+Z_1+2Z_2}}},\\
	Z_1=1+\sqrt[3]{1-a_\star^2}\br{\sqrt[3]{1+a_\star}+\sqrt[3]{1-a_\star}},\qquad
	Z_2=\pa{3a_\star^2+Z_1^2}^{1/2},\qquad
	a_\star=\frac{a}{M}.
\end{gather}
\end{subequations}
Beyond this radius, orbiters must necessarily plunge.  Following Cunningham \cite{Cunningham1975}, we consider an equatorial disk consisting of emitters on prograde circular orbits for $r_s\ge r_\mathrm{ms}$, and emitters on infalling timelike geodesics with the same conserved quantities as the ISCO for $r_+\le r_s<r_\mathrm{ms}$.  This same model was used in Ref.~\cite{GrallaLupsasca2020d} to produce the BH image in the right panel of Fig.~\ref{fig:PhotonRingShell}.  Here, we provide formulas for the observed redshift $g$ of photons received by a distant observer from such an equatorial disk.  This redshift depends only on the emission radius $r_s$ and energy-rescaled angular momentum $\lambda$ of the photon, and is given by
\begin{align}
	\label{eq:Redshift}
	g(r_s,\lambda)=
	\begin{cases}
		g_\mathrm{orbit}&r\ge r_\mathrm{ms},\\ 
		g_\mathrm{infall}&r<r_\mathrm{ms},
	\end{cases}
\end{align}
where
\begin{align}
	g_\mathrm{disk}=\frac{\sqrt{r_s^3-3Mr_s^2+2a\sqrt{M}r_s^{3/2}}}{r_s^{3/2}+\sqrt{M}\pa{a-\lambda}},\qquad
	g_\mathrm{infall}=\frac{1}{u^t-u^\phi\lambda-u^r\br{\Delta(r_s)}^{-1}\br{\pm\sqrt{\mathcal{R}(r_s)}}},
\end{align}
with $\pm=\sign\pa{p_s^r}$ corresponding to the radial direction of emission from the source, and
\begin{gather}
	u^r=-\sqrt{\frac{2}{3}\frac{M}{r_\mathrm{ms}}}\pa{\frac{r_\mathrm{ms}}{r_s}-1}^{3/2},\qquad
	u^\phi=\frac{\gamma_\mathrm{ms}}{r_s^2}\pa{\lambda_\mathrm{ms}+aH},\qquad
	u^t=\gamma_\mathrm{ms}\br{1+\frac{2M}{r_s}\pa{1+H}},\\
	H=\frac{2Mr_s-a\lambda_\mathrm{ms}}{\Delta(r_s)},\qquad
	\lambda_\mathrm{ms}=\frac{\sqrt{M}\pa{r_\mathrm{ms}^2-2a\sqrt{Mr_\mathrm{ms}}+a^2}}{r_\mathrm{ms}^{3/2}-2M\sqrt{r_\mathrm{ms}}+a\sqrt{M}},\qquad
	\gamma_\mathrm{ms}=\sqrt{1-\frac{2}{3}\frac{M}{r_\mathrm{ms}}}.
\end{gather}

\section{Corrections in this version}
\label{app:Corrections}

In this third version (v3) we have corrected a few minor errors found in previous versions. These corrections are specified below. They do not change the main conclusions of the paper. 

\begin{itemize}
    \item $(1-e^{-\gamma_0}) \to (1-e^{-2\gamma_0})$ in the denominator of Eq.~\eqref{eq:ToyModelIntermediateStep2}. We propagated this correction to all relevant subsequent equations.
    \item On the fourth and sixth lines of Eq.~\eqref{eq:SmallInclinationCorrelator},  $G_{\ell_t}\br{\ldots -\sin{\theta_o}\sin{\frac{\Phi}{2}}\pa{\ldots}} \to G_{\ell_t}\br{\ldots +\sin{\theta_o}\sin{\frac{\Phi}{2}}\pa{\ldots}}$, and $G_{\ell_\phi}^\circ\br{\ldots -\sin{\theta_o}\sin{\frac{\Phi}{2}}\pa{\ldots}} \to G_{\ell_\phi}^\circ\br{\ldots +\sin{\theta_o}\sin{\frac{\Phi}{2}}\pa{\ldots}}$, and this correction was propagated forward to all relevant equations.
    \item Added the odd-parity terms $\propto \sin \varphi$ in the small-inclination expansion, $f^o_\phi$, $f^o_t$, 
    $\iota^o_1$ in Eqs.~\eqref{eq:small inclination expansion a}, \eqref{eq:small inclination expansion b}, \eqref{eq:small inclination expansion c}, \eqref{eq:small inclination f, iota expansions} and propagated them to all relevant subsequent equations.
\end{itemize}

\bibliography{correlations.bib}
\bibliographystyle{utphys}

\end{document}